\documentclass[twocolumn,showpacs,amsmath,amssymb,aps,prc]{revtex4-1}

\RequirePackage{lineno}
\usepackage{graphicx}
\usepackage{dcolumn}
\usepackage{bm}

\newcommand{\EtaTwoG}{\eta\rightarrow 2\gamma}
\newcommand{\EtaSixG}{\eta\rightarrow 3\pi^{0}}
\newcommand{\EtaSixGFull}{\eta\rightarrow 3\pi^{0}\rightarrow 6\gamma}

\newcommand{\ReacInc}{\gamma N\rightarrow \eta (N)}

\newcommand{\ReacExcP}{\gamma p\rightarrow \eta p}
\newcommand{\ReacExcN}{\gamma n\rightarrow \eta n}

\newcommand{\ReacExcNTwoG}{\gamma n\rightarrow \eta n\rightarrow 2\gamma n}
\newcommand{\ReacExcNSixG}{\gamma n\rightarrow \eta n\rightarrow 3\pi^{0} n}

\newcommand{\ReacExcNPizPip}{\gamma p\rightarrow \pi^0\pi^+ n}

\newcommand{\SOneOne}{\mbox{$S_{11}$(1535)} }
\newcommand{\DOneThree}{\mbox{$D_{13}$(1520)} }

\newcommand{\CT}{\cos(\theta_{\eta}^{*})}
\newcommand{\CTTOF}{-1 < \cos(\theta_{\eta}^{*}) < -0.5}
\newcommand{\WK}{W_{\mathrm{kin}}}
\newcommand{\WT}{W_{\mathrm{TOF}}}

\begin{document}

\preprint{APS/123-QED}

\title{Quasifree Photoproduction of $\boldsymbol\eta$ Mesons off Protons and Neutrons}

\author{
  D.~Werthm\"uller$^1$,
  L.~Witthauer$^1$,
  I.~Keshelashvili$^{1}$, 
  P.~Aguar-Bartolom${\acute{\rm e}}$$^{2}$,   
  J.~Ahrens$^{2}$,
  J.R.M.~Annand$^{3}$,
  H.J.~Arends$^{2}$,
  K.~Bantawa$^{4}$,
  R.~Beck$^{2,5}$,
  V.~Bekrenev$^{6}$,
  A.~Braghieri$^{7}$,
  D.~Branford$^{8}$,
  W.J.~Briscoe$^{9}$,
  J.~Brudvik$^{10}$,
  S.~Cherepnya$^{11}$,
  S.~Costanza$^{7}$,
  B.~Demissie$^{9}$,
  M.~Dieterle$^{1}$,  
  E.J.~Downie$^{2,3,9}$,
  P.~Drexler$^{12}$,
  L.V. Fil'kov$^{11}$, 
  A. Fix$^{13}$,      
  D.I.~Glazier$^{8}$,
  D. Hamilton$^{3}$,
  E.~Heid$^{2}$,
  D.~Hornidge$^{14}$,
  D.~Howdle$^{3}$,
  G.M.~Huber$^{15}$,
  I.~Jaegle$^{1}$,
  O.~Jahn$^{2}$,
  T.C.~Jude$^{8}$,
  A.~K{\"a}ser$^{1}$,   
  V.L.~Kashevarov$^{2,11}$,
  R.~Kondratiev$^{16}$,
  M.~Korolija$^{17}$,
  S.P.~Kruglov$^{6}$, 
  B.~Krusche$^1$, 
  A.~Kulbardis$^{6}$,  
  V.~Lisin$^{16}$,
  K.~Livingston$^{3}$,
  I.J.D.~MacGregor$^{3}$,
  Y.~Maghrbi$^{1}$,
  J.~Mancell$^{3}$, 
  D.M.~Manley$^{4}$,
  Z.~Marinides$^{9}$,
  M.~Martinez$^{2}$,
  J.C.~McGeorge$^{3}$,
  E.F.~McNicoll$^{3}$,
  V.~Metag$^{12}$,
  D.G.~Middleton$^{14}$,
  A.~Mushkarenkov$^{7}$,
  B.M.K.~Nefkens$^{10}$,
  A.~Nikolaev$^{2,5}$,
  R.~Novotny$^{12}$,
  M.~Oberle$^{1}$,
  M.~Ostrick$^{2}$,
  P.B.~Otte$^{2}$,
  B.~Oussena$^{2,9}$, 
  P.~Pedroni$^{7}$,
  F.~Pheron$^{1}$,
  A.~Polonski$^{16}$,
  S.N.~Prakhov$^{2,9,10}$,
  J.~Robinson$^{3}$,   
  G.~Rosner$^{3}$,
  T.~Rostomyan$^{1}$,
  S.~Schumann$^{2,5}$,
  M.H.~Sikora$^{8}$,
  D.~Sober$^{18}$,
  A.~Starostin$^{10}$,
  I.~Supek$^{17}$,
  M.~Thiel$^{2,12}$,
  A.~Thomas$^{2}$,
  M.~Unverzagt$^{2,5}$,
  D.P.~Watts$^{8}$ \\
(A2 Collaboration at MAMI)
}

\affiliation{
  $^{1}$\mbox{Departement Physik, University of Basel, CH-4056 Basel, Switzerland}\\
  $^{2}$\mbox{Institut f\"ur Kernphysik, University of Mainz, D-55099, Germany}\\
  $^{3}$\mbox{SUPA, School of Physics and Astronomy, University of Glasgow, Glasgow G12 8QQ, United Kingdom}\\
  $^{4}$\mbox{Kent State University, Kent, Ohio 44242, USA}\\  
  $^{5}$\mbox{Helmholtz-Institut f\"ur Strahlen- und Kernphysik, University of Bonn, D-53115 Bonn,Germany}\\
  $^{6}$\mbox{Petersburg Nuclear Physics Institute, 188300 Gatchina, Russia}\\
  $^{7}$\mbox{INFN Sezione di Pavia, I-27100 Pavia, Italy}\\
  $^{8}$\mbox{SUPA, School of Physics, University of Edinburgh, Edinburgh EH9 3JZ, United Kingdom}\\
  $^{9}$\mbox{The George Washington University, Washington D.C. 20052, USA}\\
  $^{10}$\mbox{University of California Los Angeles, Los Angeles, California 90095-1547, USA}\\
  $^{11}$\mbox{Lebedev Physical Institute, 119991 Moscow, Russia}\\
  $^{12}$\mbox{II. Physikalisches Institut, University of Giessen, D-35392 Giessen, Germany}\\
  $^{13}$\mbox{Laboratory of Mathematical Physics, Tomsk Polytechnic University, 634050 Tomsk, Russia}\\
  $^{14}$\mbox{Mount Allison University, Sackville, New Brunswick E4L 1E6, Canada}\\
  $^{15}$\mbox{University of Regina, Regina, Saskatchewan S4S 0A2, Canada}\\
  $^{16}$\mbox{Institute for Nuclear Research, 125047 Moscow, Russia}\\
  $^{17}$\mbox{Rudjer Boskovic Institute, HR-10000 Zagreb, Croatia}\\
  $^{18}$\mbox{The Catholic University of America, Washington D.C. 20064, USA}\\
}


\date{\today}

\begin{abstract}
Differential and total cross sections for the quasifree reactions $\ReacExcP$ and 
$\ReacExcN$ have been determined at the MAMI-C electron accelerator 
using a liquid deuterium target. Photons were produced via bremsstrahlung from the 1.5 GeV 
incident electron beam and energy-tagged with the Glasgow photon tagger. Decay photons of 
the neutral decay modes $\EtaTwoG$ and $\EtaSixGFull$ and coincident recoil 
nucleons were detected in a combined setup of the Crystal Ball and the TAPS calorimeters. 
The $\eta$-production cross sections were measured in coincidence with recoil protons,
recoil neutrons, and in an inclusive mode without a condition on recoil nucleons,
which allowed a check of the internal consistency of the data. The effects from nuclear
Fermi motion were removed by a kinematic reconstruction of the final-state invariant mass
and possible nuclear effects on the quasifree cross section were investigated by a
comparison of free and quasifree proton data. The results, which represent a significant 
improvement in statistical quality compared to previous measurements, agree with the known 
neutron-to-proton cross-section ratio in the peak of the \SOneOne resonance and confirm 
a peak in the neutron cross section, which is absent for the proton, at a center-of-mass 
energy $W = (1670\pm 5)$ MeV with an intrinsic width of $\Gamma\approx 30$~MeV.
\end{abstract}


\maketitle

\section{Introduction}
Photoproduction of mesons provides important information about the excitation spectrum of 
the nucleon that, despite various long-lasting experimental and theoretical efforts, is 
still not sufficiently understood. The number of predicted states (see Review of Quark Model
in Ref.~\cite{PDG_12} exceeds the experimentally observed number and the properties of 
some identified states are markedly different from those expected.
The difficulty of identifying the relevant degrees of freedom of Quantum Chromodynamics (QCD)
in the nonperturbative region using effective models can perhaps be solved in the future 
by lattice calculations. Nevertheless, further precise experimental input is needed since 
the majority of the available data comes from pion scattering experiments, 
which could leave states that couple only weakly to $\pi N$ undiscovered. This situation is
currently changing due to the world-wide effort to measure the photoproduction of mesons
off nucleons with tagged photon beams. Not only angular distributions but also many different
polarization observables from measurements with circularly and linearly polarized beams
and longitudinally and transversely polarized targets are becoming available, e.g., from 
the CLAS experiment at the Thomas Jefferson National Accelerator Facility (JLab),
the CBELSA/TAPS experiment at ELSA, and the Crystal Ball/TAPS
experiment at MAMI. The first results of this program are summarized in the Baryon Particle Listings
in the Review of Particle Physics \cite{PDG_12} by the Particle Data Group (PDG). The new results 
from photon-induced reactions are quite important, e.g., for the coupled-channel partial-wave 
analysis of the Bonn-Gatchina group (BnGn) \cite{Anisovich_12} or the partial-wave analysis of
the SAID group \cite{Workmann_12}.

The majority of these recent measurements investigated photoproduction off free protons. 
The complementary program for the neutron target is less advanced due to the complications 
arising from the use of quasifree neutrons bound in light nuclei as targets. However, this 
program is very important for the investigation of $N^{\star}$ resonances because the 
$\gamma NN^{\star}$ helicity couplings are isospin dependent. In some cases it is even 
possible that the excitation of states is forbidden for the proton (or at least strongly 
suppressed) but allowed for the neutron (Moorehouse selection rules \cite{Moorehouse_66}). 
Therefore, the isospin decomposition of the amplitudes requires measurements
of photoproduction of mesons off neutrons. Light nuclei such as $^3$He and, in particular, 
the loosely bound deuteron are the best available targets. In comparison to measurements 
with free protons some complications arise. The first is of a technical nature. 
The classification of the final state requires the detection and identification of the 
recoil nucleon. This is challenging for all-neutral final states (neutron, decay-photons 
from neutral mesons like $\pi^0$, $\eta$), which are produced in some of the most interesting reactions. 
At present, only almost $4\pi$ covering electromagnetic calorimeters with good particle 
identification capabilities can efficiently measure such reactions. All excitation functions, 
angular distributions, and other observables will be smeared by the Fermi motion of the 
bound nucleons. This problem can in principle be overcome by a complete reconstruction of 
the final-state nucleon-meson kinematics, which reveals the `true' 
$W = \sqrt{s} = \sqrt{(p_{N'}+p_m)^2}$ ($p_{N'}$, $p_m$: recoil nucleon and meson four-momenta)
of the reaction. However, for tagged-photon experiments this 
means that the resolution of $W$ is no longer defined by the energy resolution 
of the incident tagged-photon beam but by the typical resolution of the reaction-product 
detector (which is usually much worse). The last problem is the possible modification of the 
experimental results due to nuclear effects, in particular final-state interactions (FSI)
between the nucleons or between mesons and nucleons. 
Such effects can be investigated for proton photoreactions where the results for free protons
can be compared with quasifree measurements on protons bound in the deuteron. This gives some 
indication of whether, for a specific reaction channel, FSI effects are important and can 
test FSI models before they are applied to quasifree neutron measurements.

\subsection{Photoproduction of $\eta$ mesons} 
Photoproduction of $\eta$ mesons attracted much interest when this reaction became 
experimentally accessible with a precision comparable to pion photoproduction. Due to its
isoscalar nature only isospin 1/2 $N^{\star}$ resonances can decay to the nucleon ground state 
via $\eta$ emission. Furthermore, due to its relatively large mass (compared to pions) the number
of relevant partial waves is still small at excitation energies where so far many predicted 
low-spin $N^{\star}$ states are `missing'. This simplifies the interpretation of the data. 

\begin{figure}[b]
\centering
\includegraphics[width=0.48\textwidth]{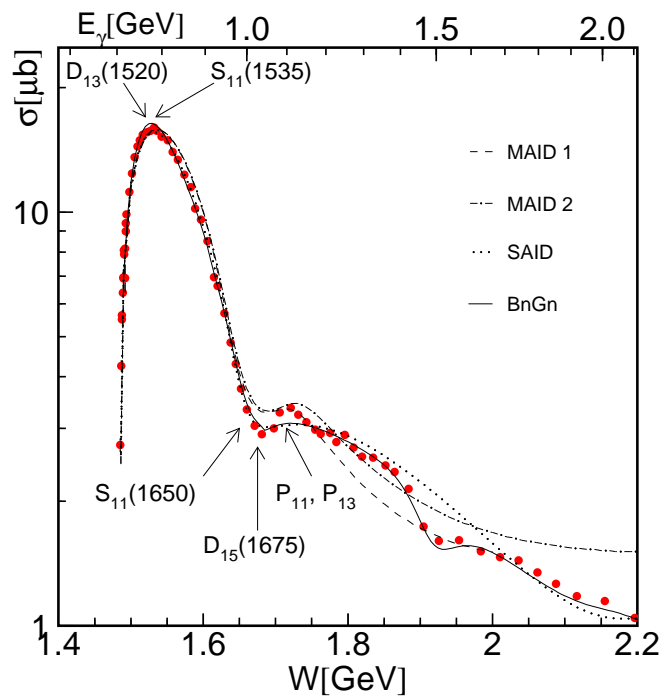}
\caption{(Color online) Total cross section for $\ReacExcP$ averaged over data from 
\cite{Krusche_95,Renard_02,Dugger_02,Crede_05,Nakabayashi_06,Bartalini_07,Crede_09,Williams_09,McNicoll_10}. 
Model curves are from MAID 1 \cite{Chiang_02}, MAID 2 \cite{Chiang_03}, SAID \cite{McNicoll_10}, and 
BnGn \cite{Anisovich_05}.}
\label{fig:proton}
\end{figure}

Experimental progress for measurements of the $\gamma p\rightarrow p\eta$ reaction
with free protons was huge during the last decade, so that now photoproduction of 
$\eta$ mesons is probably the best studied final state apart from pion 
production. Total cross sections, angular distributions, and some polarization observables have been 
investigated at all major tagged photon facilities sometimes even with repeated and improved 
experiments \cite{Krusche_95,Ajaka_98,Bock_98,Renard_02,Dugger_02,Crede_05,Nakabayashi_06,Bartalini_07,Bartholomy_07,Elsner_07,Crede_09,Williams_09,Sumihama_09,McNicoll_10}.  
For the discussion of the gross features of $\eta$ production off the proton, Fig.~\ref{fig:proton} 
summarizes the results for the total cross section (averaged over all available data). 

At threshold ($E_{\gamma}$= 708~MeV, $W=$ 1486~MeV) the reaction is completely dominated by the 
excitation of the $S_{11}$(1535) resonance \cite{Krusche_97}. Contributions from the $P_{11}$(1440)
(Roper) resonance have never been directly identified. The Review of Particle Physics cites a
branching ratio of (0$\pm$1)\% \cite{PDG_12}. The $D_{13}$(1520) resonance makes a tiny contribution 
(branching ratio (0.23$\pm$0.04)\% \cite{PDG_12}), which is negligible for the total cross section
but was identified via an interference term in the angular distributions \cite{Krusche_95,Weiss_03}
and, even more significantly, in the photon beam asymmetry \cite{Ajaka_98,Elsner_07}. At slightly
higher energy the $S_{11}$(1650) interferes (for the proton) destructively with the $S_{11}$(1535).
In the S$_{11}$ region (see \cite{Krusche_03} for a summary) contributions from
non-resonant backgrounds seem to be small. 

At slightly higher energies contributions from the $D_{15}$(1675), $D_{13}$(1700), $P_{11}$(1710),
and $P_{13}$(1720) resonances can be expected. Branching ratio estimates given by PDG \cite{PDG_12}
are 10--30\% for the $P_{11}$, (4$\pm$1)\% for the $P_{13}$, and (0$\pm$1)\% for the two $d$-wave
states. The results differ between the available analyses. Total cross sections from some analyses
(MAID 1: $\eta$-MAID isobar model \cite{Chiang_02}, MAID 2: $\eta$-MAID reggeized isobar model
\cite{Chiang_03}, BnGn: Bonn-Gatchina coupled-channel analysis \cite{Anisovich_05},
SAID partial-wave analysis \cite{McNicoll_10}) are compared in Fig.~\ref{fig:proton} to the average
of all available data. In the region around photon energies of 1~GeV agreement between
the model fits is not very good and, as discussed, e.g., in \cite{Elsner_07}, the relative contributions
of different resonances differ quite a bit in the models.  

For the neutron, the range from threshold throughout the $S_{11}$(1535) resonance has been studied 
intensively using the deuteron or helium isotopes as targets 
\cite{Krusche_95_2,Hoffmann_Rothe_97,Weiss_03,Weiss_01,Pfeiffer_04,Hejny_99,Hejny_02}. 
The experiments found consistently a neutron-to-proton cross-section ratio for quasifree production
close to 2/3 and very small coherent contributions, which was interpreted as evidence for a dominant
isovector excitation of the $S_{11}$(1535) (the  isoscalar admixture in the proton
amplitude is only $\approx$9\% \cite{Krusche_03}). 

Above this energy range many open questions exist. Most analyses (MAID, SAID) find a negative
sign for the $A^n_{1/2}$ (in the following all values are in units of $10^{-3}$ GeV$^{-1/2}$)
neutron helicity coupling of the four star $S_{11}$(1650) resonance (PDG: $-$15$\pm$21) and thus 
a destructive interference between the two $S_{11}$ states. A negative sign is also preferred by 
quark models (e.g., Capstick \cite{Capstick_92}: $-35$; Burkert et al. \cite{Burkert_03}: $-$31$\pm$3). 
However, the more recent analyses of the Bonn-Gatchina group \cite{Anisovich_13} (25$\pm$20) 
and Shresta and Manley \cite{Shrestha_12} (11$\pm$2) found positive values corresponding 
to a constructive interference. The $\eta$-MAID model \cite{Chiang_02} found a much larger 
$\eta$-decay branching ratio than quoted in PDG for the $D_{15}$(1675) state (17\%). It thus 
predicted a significant contribution of this state to $\gamma n\rightarrow n\eta$ because it
has much larger photon couplings for the neutron than for the proton. Furthermore, there were
predictions that the non-strange $P_{11}$-like member of the conjectured baryon antidecuplet 
\cite{Diakonov_97} should be electromagnetically excited more strongly on the neutron, should 
have a large decay branching ratio to $N\eta$, an invariant mass around 1.7 GeV, and a width 
of a few tens of MeV \cite{Diakonov_97,Polyakov_03,Arndt_04}.
  
Motivated by these open problems several experiments have recently studied this reaction.   
Exclusive measurements of $\ReacExcP$ and $\ReacExcN$ on the deuteron in quasifree 
kinematics were performed by GRAAL \cite{Kuznetsov_07}, at the Laboratory of Nuclear Science
at Sendai (LNS-Sendai) \cite{Miyahara_07} 
and by the CBELSA/TAPS collaboration \cite{Jaegle_08,Jaegle_11}. A prominent structure
in the total cross section of $\gamma n\rightarrow n\eta$ at incident photon energies
around 1~GeV was first found by the GRAAL experiment. This peak with a width of only $\approx$25~MeV 
appeared also in the CBELSA/TAPS data at $W = 1.67$ GeV when the true center-of-mass energy 
$W$ was reconstructed from the final-state $\eta$ meson and the recoil neutron. Using the 
same analysis, the cross section for $\ReacExcP$ was determined and good agreement to 
previous direct measurements on the free proton was found, demonstrating that nuclear 
effects could be reliably controlled by this method. Moreover, it was found that around the 
same value of $W$ the total proton cross section shows a dip-like structure, which was 
confirmed by the latest high-statistics measurement on free protons at MAMI-C \cite{McNicoll_10}.
The origin of the dip in the proton cross section was recently discussed for various scenarios
(narrow resonances, threshold effects from the $\gamma p\rightarrow \omega p$ reaction)
in the framework of the BnGn model \cite{Anisovich_13a}.  

At the moment the nature of the peak in the neutron cross section and the dip in the proton 
cross section is not understood nor is it clear whether they are correlated. The only
scenario that is ruled out in the case of the neutron is that the peak originates from an 
isolated conventional broad resonance. Various scenarios have been suggested in the literature. 
They range from different coupled-channel effects of known nucleon resonances 
\cite{Shklyar_07,Shyam_08}, interference effects in the $S_{11}$ partial wave 
\cite{Anisovich_09, Anisovich_13}, effects from strangeness threshold openings 
\cite{Doering_10}, to intrinsically narrow states 
\cite{Arndt_04,Choi_06,Fix_07,Anisovich_09,Shrestha_12}. 
The available data are insufficient for an unambiguous analysis.

In this work we present in detail the results from a new high statistics measurement of the
total cross section and angular distributions for the reactions $\ReacExcP$ and $\ReacExcN$ 
extracted from data taken with a liquid deuterium target at Mainz.
The main experimental results have been summarized in a preceding Letter \cite{Werthmueller_13}.
Both the $\EtaTwoG$ and the $\EtaSixGFull$ neutral decay modes were used for the reconstruction 
of the $\eta$ meson, leading to an unprecedented statistical quality of the results and to 
stringent limits for systematic uncertainties.

Measurements of further quantities, such as single and double polarization observables, are 
of course highly desirable and are already under way at MAMI. 

\section{Experimental setup}
The data were measured during three different beam times at the MAMI electron accelerator 
facility in Mainz \cite{Herminghaus_76,Kaiser_08} using the standard A2 setup for photon 
beam experiments. Details for the experimental parameters (targets, beams) are summarized 
in \cite{Oberle_13}. The electron beam, having an energy of 
$E_0 = 1508$~MeV (1558~MeV during part of the beam time) and a current ranging from 4.5 
to 10 nA after the last accelerator stage of MAMI-C, was used to produce photons via 
bremsstrahlung in a 10 $\mu$m copper radiator. The scattered electrons were momentum-analyzed 
up to 95\% of the initial beam energy in the Glasgow photon tagger 
\cite{Anthony_91,Hall_96,McGeorge_08}. The large magnetic dipole of the spectrometer and 
the 353 half-overlapping plastic scintillators installed in the focal plane allow an energy
reconstruction of the bremsstrahlung photons via $E_\gamma = E_0 - E_{e^-}$ from the 
measured electron energy $E_{e^-}$ with a resolution of 2--5 MeV. Electrons corresponding 
to photon energies below 400 MeV were not recorded to increase the event rate and to prevent
damage to the phototubes of the low energy tagger detectors.

\begin{figure}[t]
\centering
\includegraphics[width=0.48\textwidth]{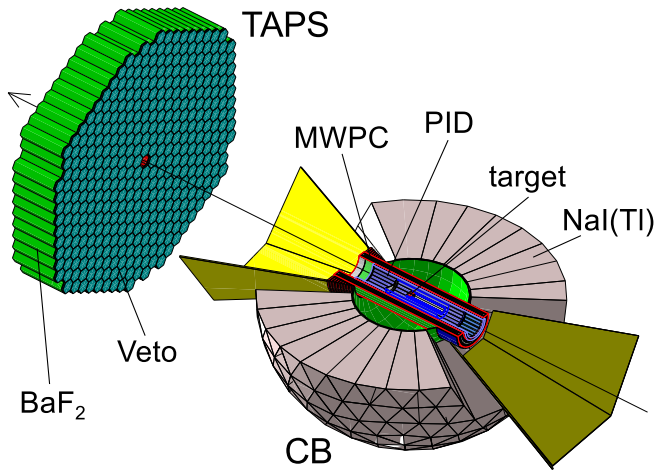}
\caption{(Color online) Diagram of the experimental setup created by the Geant4 model. The Crystal 
Ball detector was cut along the x-axis to show the inner detectors and the target.}
\label{fig:setup}
\end{figure}

The photon beam was collimated using a 4 mm diameter lead collimator and impinged on the 4.72 cm 
long (3.0 cm for part of the beam time) Kapton target cylinder of 4 cm diameter. Outgoing 
particles were detected using the two calorimeters Crystal Ball (CB) \cite{Starostin_01} 
and TAPS \cite{Novotny_91,Gabler_94}. A schematic diagram of the detector setup is 
shown in Fig.~\ref{fig:setup}. The CB consists of two hemispheres with in total 672 
optically insulated NaI(Tl) crystals of 15.7 radiation length thickness, covering all 
azimuthal angles for the polar angle range $20^\circ < \theta < 160^\circ$. 
All crystals point towards the center of the sphere where the target is mounted.
The distance from the center to the detector modules is 25 cm.
The energy resolution for photons can be described as 
$\Delta E / E = 2\%/(E[\mathrm{GeV}])^{0.36}$ while typical angular resolutions are 
$\Delta\theta = 2^\circ$--$3^\circ$ and $\Delta\phi = 2^\circ$--$4^\circ$ \cite{Starostin_01}.

The forward hole of the CB is covered by the hexagonal TAPS wall, which is made of 
384 hexagonally shaped BaF$_2$ crystals with a thickness of 12 radiation lengths. TAPS was 
installed 1.46 m downstream from the target covering polar angles from 
$5^\circ$ to $21^\circ$. The photon energy resolution is parametrized as 
$\Delta E / E = 1.8\% + 0.8\%/(E[\mathrm{GeV}])^{0.5}$ \cite{Gabler_94}. 
The fine granularity of the detector
elements leads to excellent resolution in the polar angle (better than $1^\circ$), while 
$\Delta\phi = 1^\circ$--$6^\circ$.

Neutral and charged particles were discriminated by plastic scintillators in both detectors.
A 50 cm long barrel of 24 strips with a width of 4 mm surrounded the target and acted as 
particle identification detector (PID) for the CB \cite{WATTS_05}. In TAPS charged particles 
were identified with individual 5 mm thick plastic scintillators that were installed in front 
of every detector element. The multi-wire proportional chamber (MWPC) surrounding the PID 
in the CB was not active for the present experiment.

The experimental trigger was relatively open and consisted of a total energy sum threshold 
in the CB and a minimal total `hit' multiplicity in the CB and TAPS. The energy sum was implemented 
as the analog sum of all signals from the CB and its threshold was adjusted to correspond to 
an energy of around 300 MeV, mainly to reject $\pi^0$ production events. 
The 672 crystals of the CB were grouped into 45 sectors each containing up to 16
neighboring crystals and TAPS was divided into 6 triangular sectors. If the signal from at least one crystal
in a sector exceeded a threshold of about 30 MeV (35 MeV in TAPS) that sector contributed to the event
multiplicity.
A minimal total multiplicity in the CB and TAPS 
of two was set to accept events from the $\EtaTwoG$ decay. For a part of the beam time
a multiplicity of 3 was required to increase statistics for the $\EtaSixG$ decay and other
multi-photon channels. This data set could not be used for the $\EtaTwoG$ decay.

\section{Data analysis}

\subsection{Subtraction of tagger random coincidences}

\begin{figure}[th]
\centering
\includegraphics[width=0.48\textwidth]{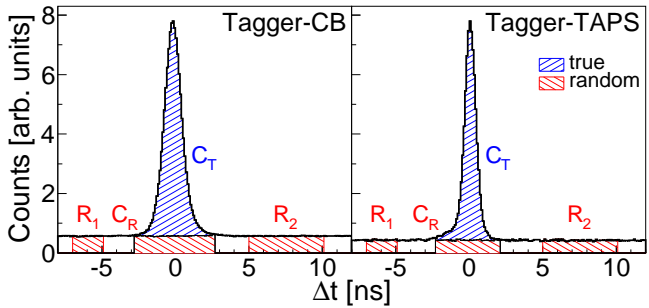}
\caption{(Color online) Tagger-calorimeter coincidence time spectra (sum of all channels): Left side: timing between tagger 
and the Crystal Ball. Right side: timing between tagger and TAPS. 
Hatched red areas: random background ($R_1$ and $R_2$) and random coincidence ($C_R$) 
windows. Hatched blue areas: true coincidence window ($C_T$).}
\label{fig:rnd_subtr}
\end{figure}

All electron hits in the photon tagger were recorded for each event triggered by the 
calorimeters. The tagger itself did not contribute to the trigger decision because 
for every CB/TAPS trigger there was almost always a hit in the tagger.
The random coincidences were subtracted by a standard side-band analysis of the production-detector
tagger coincidence. Fig.~\ref{fig:rnd_subtr} shows the relative timing spectra between the 
tagger and both the CB and TAPS detectors. The time resolution with respect to TAPS was around 
0.9 ns compared to the 1.5 ns that could be achieved with the CB. 
Hence, whenever possible, tagger coincidence time evaluation was performed with photons 
detected in TAPS. The true coincidences located in the peak ($C_T$) (hatched blue area in 
Fig.~\ref{fig:rnd_subtr}) were determined by a subtraction of the random coincidences ($C_R$), 
which were analyzed with cuts on the random background in the time spectrum (regions $R_1$, $R_2$) 
with proper normalization. This procedure was applied to all spectra. The random background
windows in Fig.~\ref{fig:rnd_subtr} are only a schematic representation. They were much wider
in the actual analysis ($2\times 200$ ns) so that statistical uncertainties from random 
background were negligible.

\begin{table}[b]
\centering
\begin{tabular}{|c|c|c|}
\hline $\eta$ decay & reaction & cluster selection criteria \\
\hline 
& $\ReacInc$ & (2n \& 0c) or (2n \& 1c) or (3n \& 0c) \\
$2\gamma$ & $\ReacExcP$ & 2n \& 1c \\
& $\ReacExcN$ & 3n \& 0c \\
\hline
& $\ReacInc$ & (6n \& 0c) or (6n \& 1c) or (7n \& 0c) \\
$3\pi^0$ & $\ReacExcP$ & 6n \& 1c \\
& $\ReacExcN$ & 7n \& 0c \\ \hline
\end{tabular}
\caption{Summary of the basic event selection criteria based on the number and type 
(n=neutral, c=charged) of detected clusters. See text for more details.}
\label{table:event_selection}
\end{table}

\subsection{Particle reconstruction and reaction identification}
\label{sec:iden}

Clusters in the calorimeters were built from adjacent crystals where the deposited energy 
in each single crystal exceeded a typical threshold of 2 MeV in the CB and 3--5 MeV in TAPS. 
If the total deposited energy in all crystals was below 20 MeV, the cluster was ignored in 
the analysis.

The reconstructed clusters in the CB and TAPS were first classified as neutral or charged.
Clusters were marked as charged if a coincident hit in the corresponding PID (CB) or 
veto element (TAPS) was found, otherwise they were marked as neutral. Depending on the 
reaction to be measured, a condition on the number and type of clusters was set. 
An overview of these conditions is given in Table \ref{table:event_selection}. To measure 
the reactions $\ReacExcP$ and $\ReacExcN$ exclusive measurements were performed, i.e., the 
detection of the recoil nucleons was required. In the inclusive measurement of $\ReacInc$ 
the recoil nucleon could also be undetected. The $\eta$ meson was identified using the 
neutral decays $\EtaTwoG$ and $\EtaSixGFull$ as described in the following.

\begin{figure}[t]
\centering
\includegraphics[width=0.5\textwidth]{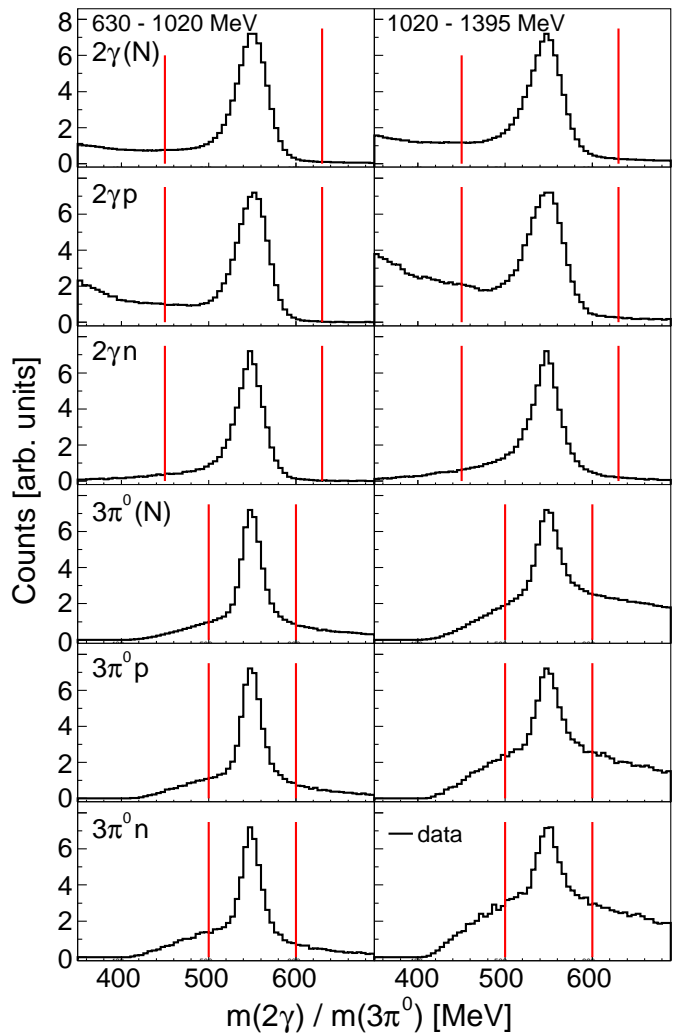}
\caption{(Color online) Typical invariant-mass spectra for two bins of incident photon energy $E_\gamma$ 
in the range of $\eta$ mesons for the $\EtaTwoG$
and $\EtaSixG$ decays in coincidence with recoil protons $p$, recoil neutrons $n$,
and without any condition for recoil nucleons $(N)$. The indicated cuts (red lines) have been
applied to the spectra discussed below.
}
\label{fig:inv_mass_raw}
\end{figure}

\begin{figure*}[t]
\centering
\includegraphics[width=0.99\textwidth]{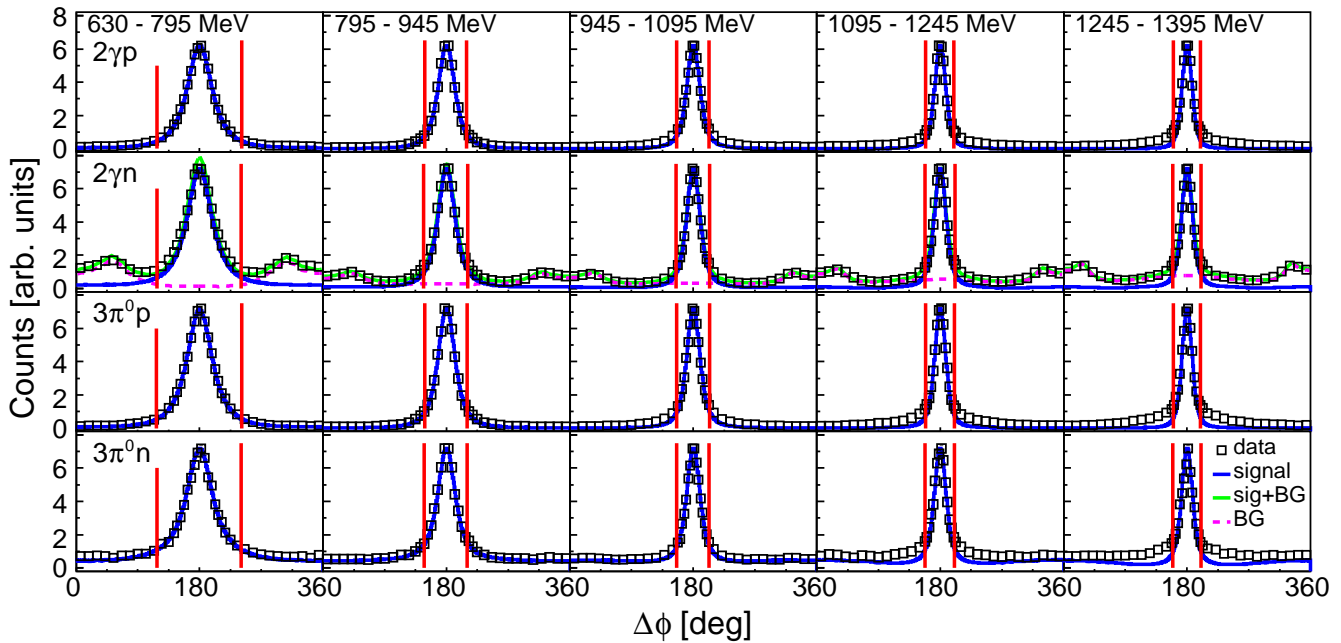}
\caption{(Color online) $\eta$-$N$ coplanarity $\Delta\phi$ used for the reaction identification: 
Top two rows: $\EtaTwoG$ analyses. Bottom two rows: $\EtaSixG$ analyses. 
Columns: bins of incident photon energy $E_\gamma$. Black squares: experimental data. 
Curves: simulations of pure signal (blue), all background contributions (dashed magenta, 
see text), sum of signal and background (green). Red lines: cut markers.}
\label{fig:cop}
\end{figure*}

\begin{figure*}[t]
\centering
\includegraphics[width=0.99\textwidth]{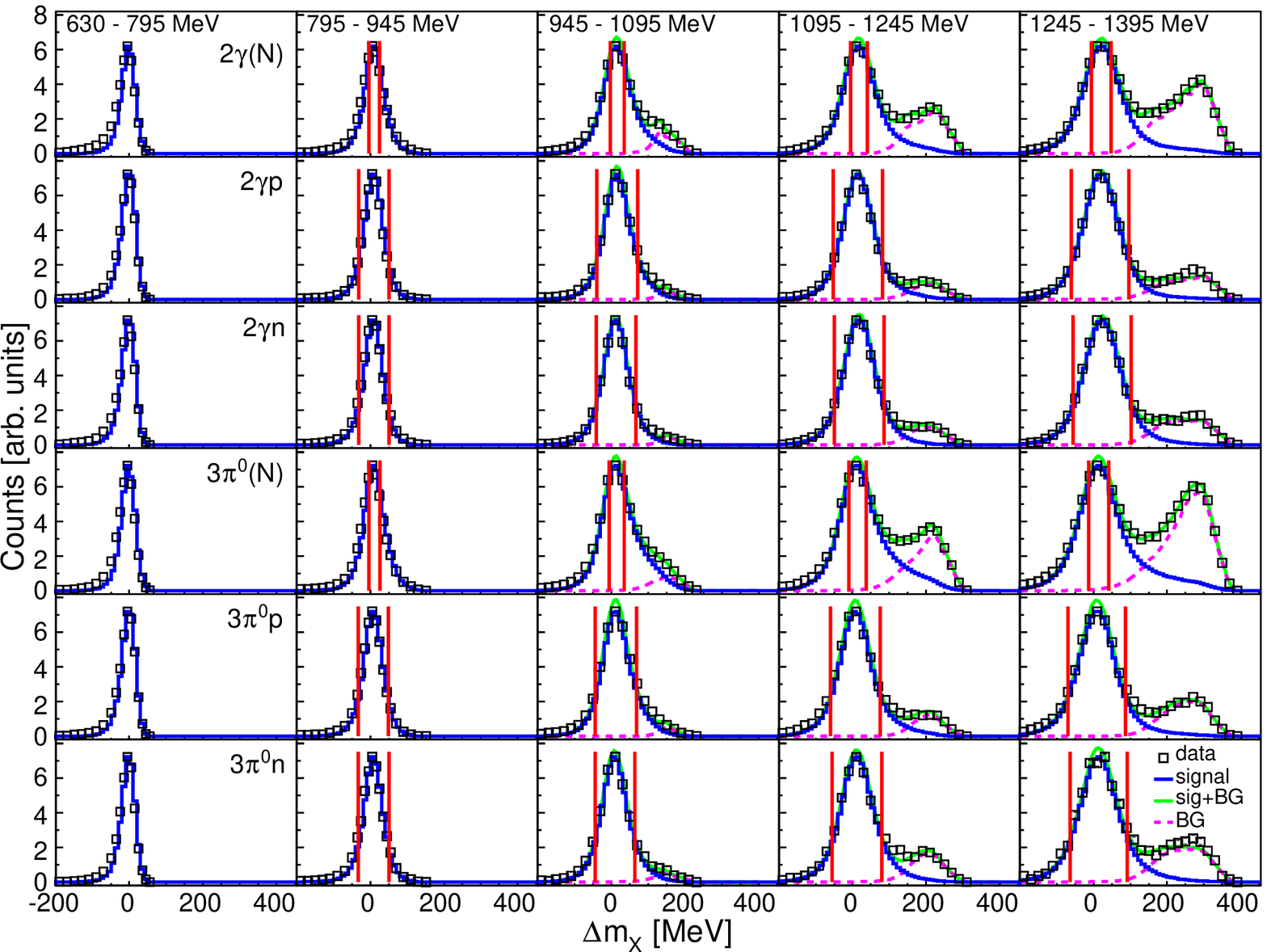}
\caption{(Color online) Missing mass $\Delta m_X$ used for the reaction identification: 
Top three rows: $\EtaTwoG$ analyses. Bottom three rows: $\EtaSixG$ analyses. 
Columns: bins of incident photon energy $E_\gamma$. Black squares: experimental data. 
Curves: simulations of pure signal (blue), all background contributions (dashed magenta, 
see text), sum of signal and background (green). Red lines: cut markers.}
\label{fig:mm}
\end{figure*}

The charged cluster was always assumed to originate from the recoil proton. In the 
$\EtaTwoG$ analysis with proton coincidence the $\eta$ meson four-momentum was then immediately 
reconstructed from the remaining two neutral clusters. In the case of neutron coincidence, 
a $\chi^2$ search was performed among the detected three neutral clusters finding the 
minimal
\begin{equation}
\chi^2 = \left( \frac{m_\eta - m_{\gamma\gamma}}{\Delta m_{\gamma\gamma}}\right)^2
\end{equation}
of all three photon pair combinations. The invariant mass of the two photons and 
the real $\eta$ mass are denoted as $m_{\gamma\gamma}$ and $m_\eta$, respectively. 
The uncertainty of the invariant mass $\Delta m_{\gamma\gamma}$ was evaluated with respect to 
individual photon cluster angular and energy resolutions of the detector system. The required
resolutions $\Delta\theta(\theta), \Delta\phi(\theta)$ and $\Delta E(E)$ were determined 
from Monte Carlo (MC) simulations and the known energy resolutions of the CB and TAPS.
Once the best combination was found the remaining cluster was marked as the neutron candidate. 
Events with wrong assignments of the neutrons are sufficiently rejected by the later
applied analysis cuts (see Sec.~\ref{sec:add_checks}).
In the $\EtaSixG$ analyses the minimal 
\begin{equation}
\chi^2 = \sum\limits_{i=1}^{3}\left( \frac{m_{\pi^0} - m_{\gamma\gamma,i}}{\Delta m_{\gamma\gamma,i}}\right)^2
\end{equation}
over all possible combinations to form three $\pi^0$ mesons with masses $m_{\pi^0}$ out of 
six or seven neutral clusters was used to combine the best three $\pi^0$ mesons to 
an $\eta$ meson. More details about the $\chi^2$ analysis are given in \cite{Witthauer_13}.

Since the contribution of the energy resolution to the two photon invariant mass is larger
than the contribution of the angular resolution, the
energy reconstruction of mesons can be optimized by applying the correction
\begin{equation}
E^{'} = E \frac{m_m}{m_{\gamma\gamma}}
\end{equation}
to the reconstructed energy $E$, where $m_{\gamma\gamma}$ and $m_m$ are the invariant mass 
of the decay photons and the real meson mass, respectively, thus obtaining the energy 
$E^{'}$, which has better resolution. This method correcting the detected photon energies
within their resolution was used for the final $\eta$ reconstruction as well as for the 
intermediate state $\pi^0$ reconstruction in the $\EtaSixG$ analyses.

Typical spectra of the $2\gamma$ and $3\pi^0$ invariant masses obtained after this event 
selection are shown in Fig.~\ref{fig:inv_mass_raw} for two ranges of incident photon energy. 
The resolution in the $\EtaSixG$ channel is better due to the constraints posed by the 
intermediate state $\pi^0$ mesons (cuts of $\pm3\sigma$ were applied on the $m_{\pi^0}(2\gamma)$ 
masses in the reconstruction of the three intermediate $\pi^0$ mesons (not shown)). For all 
further spectra aiming at the reaction identification, rough cuts on the invariant 
$\eta$-mass indicated in Fig.~\ref{fig:inv_mass_raw} were applied to suppress 
backgrounds from single and double pion production.

The major part of the background from other reaction channels was removed by a 
proper identification of the signal reaction using kinematic cuts. These cuts  
were already applied before the use of further particle identification spectra, such as pulse-shape 
analysis, time-of-flight versus energy, and $\Delta E -E$ analysis, because they can be much more 
reliably modeled by MC simulations and are therefore less critical sources for 
systematic uncertainties.

For the exclusive analyses (which are more important) a coplanarity cut involving the detected 
recoil nucleons can be established. 
Namely, it is required that the $\eta$ meson, the recoil nucleon, and the incoming photon lie 
in one plane. This can be translated into a condition on the azimuthal 
angle difference $\Delta\phi$ between $\eta$ meson and recoil nucleon $N$ using
\begin{equation}
\Delta\phi = 
\begin{cases} 
\phi_\eta - \phi_N &\mbox{if } \phi_\eta - \phi_N \geq 0 \\
2\pi - |\phi_\eta - \phi_N| &\mbox{if } \phi_\eta - \phi_N < 0 
\end{cases}
,
\end{equation}
where $\phi_i$ are the corresponding azimuthal angles of the two reconstructed particles in 
the lab frame. The resulting distributions are shown in Fig.~\ref{fig:cop} along with the applied
cuts at $\pm 2\sigma$. Because of the 
Fermi motion of the initial-state nucleons, the distributions are broader than in the 
analysis of free proton data. The distributions are very well reproduced by MC
simulations. Significant background is only visible for the $2\gamma n$ final state. It originates
from the $\pi^0 n$ final state when one of the decay photons is misidentified as a neutron and
the corresponding neutron assigned as a photon. Background reactions where all final-state
particles have been detected and correctly identified will of course pass this cut. 

More sensitive is a cut on the missing mass $\Delta m_X$ of the reaction, which was calculated under 
the assumption of quasifree production of $\eta$ mesons off nucleons at rest via 
\begin{equation}
\Delta m_{X} = \sqrt{\left(E_\gamma + m_N - E_\eta\right)^2 - \left(\vec{p}_\gamma - \vec{p}_\eta\right)^2}-m_N,
\end{equation}
where $E_\gamma$, $\vec{p}_{\gamma}$ are respectively the energy and momentum of the incident photon,
$E_\eta$, $\vec{p}_\eta$ are respectively the reconstructed energy and momentum of the meson and $m_N$ is 
the mass of the recoil nucleon. The latter (no matter if detected or not) was treated 
as a missing particle. Effects from nuclear Fermi motion, which were ignored in this analysis,
broaden the experimentally observed structures. Typical spectra for $\EtaTwoG$ and $\EtaSixG$ decays in
inclusive mode and in coincidence with recoil nucleons are summarized in Fig.~\ref{fig:mm}.
The reaction identification cuts discussed above were applied to these spectra.
Events from quasifree $\eta$-production peak around zero missing mass, while background
from $\eta\pi$ final states, where the pion escaped detection, appears at positive missing 
mass and increases strongly with incident-photon energy. Because of this energy dependence, 
$E_\gamma$-dependent symmetric cuts around the signal maximum were used. They account also for the  
small offsets (identical for data and MC simulations) of the peak positions from zero for higher 
incident photon energies.

The background from $\eta\pi$ production (mainly charged pions) can pass all previous cuts when
the charged pion is emitted with low energy or at small polar angle and escapes detection.
In that case, the coplanarity cut does not help because either the energy of 
the pion is so low or its polar angle is so small that it does not disturb the azimuthal 
correlation between recoil nucleon and $\eta$ meson. Nevertheless, the coplanarity cut
removes a significant fraction of the background as can be seen (Fig.~\ref{fig:mm}) in the 
comparison of the  missing mass spectra for the exclusive reactions (with coplanarity cut) 
to the inclusive reaction (without it). As a consequence, a very strict $\pm0.5\sigma$ 
missing-mass cut was applied to the inclusive reactions whereas in the exclusive cases a 
broader $\pm1.5\sigma$ cut was applied. For photon energies below the $\eta\pi$ production threshold 
no cut was applied because no background was visible. Sufficient rejection of the background 
was checked using simulations of various $\eta\pi$ production channels (their relative
cross sections are known) that gave together with the simulated signal distributions
a good description of the measured distributions (see green curves).

\begin{figure}[t]
\centering
\includegraphics[width=0.48\textwidth]{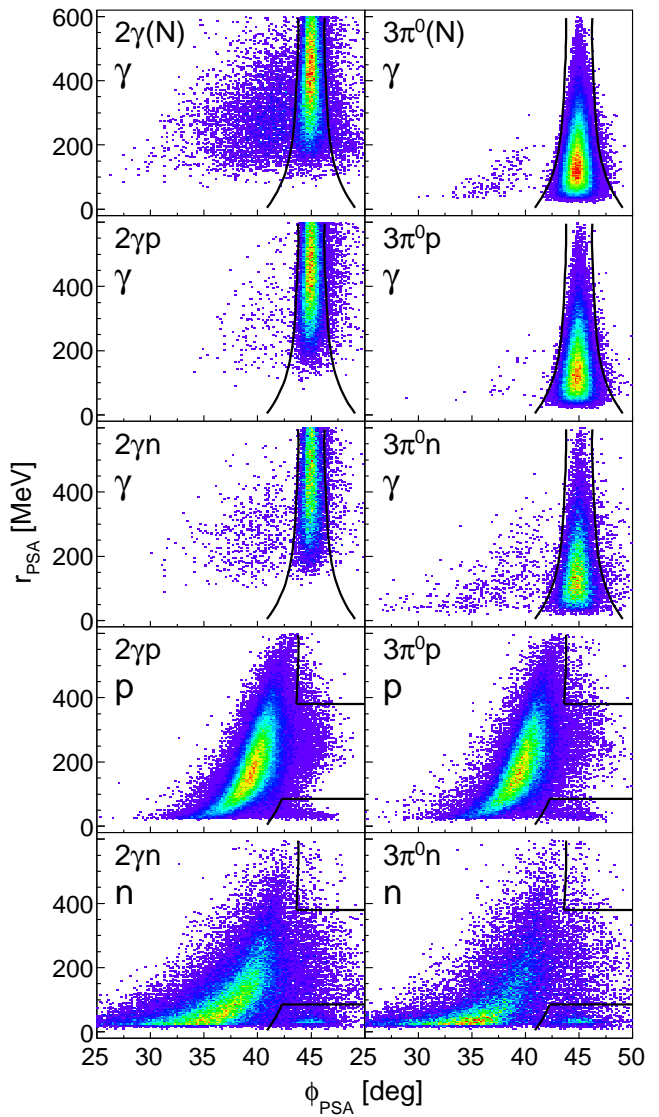}
\caption{(Color online) Pulse-shape analysis (PSA) distributions for particles detected in TAPS: 
Left column: $\EtaTwoG$ analyses. Right column: $\EtaSixG$ analyses. 
Top three rows: photon candidates in the inclusive and the two exclusive analyses. 
Bottom two rows: proton and neutron candidates in the exclusive analyses. 
Black lines: cut markers. Counts increase from violet to red.}
\label{fig:psa}
\end{figure}

\subsection{Additional checks}
\label{sec:add_checks}

\begin{figure*}[t]
\centering
\includegraphics[width=0.47\textwidth]{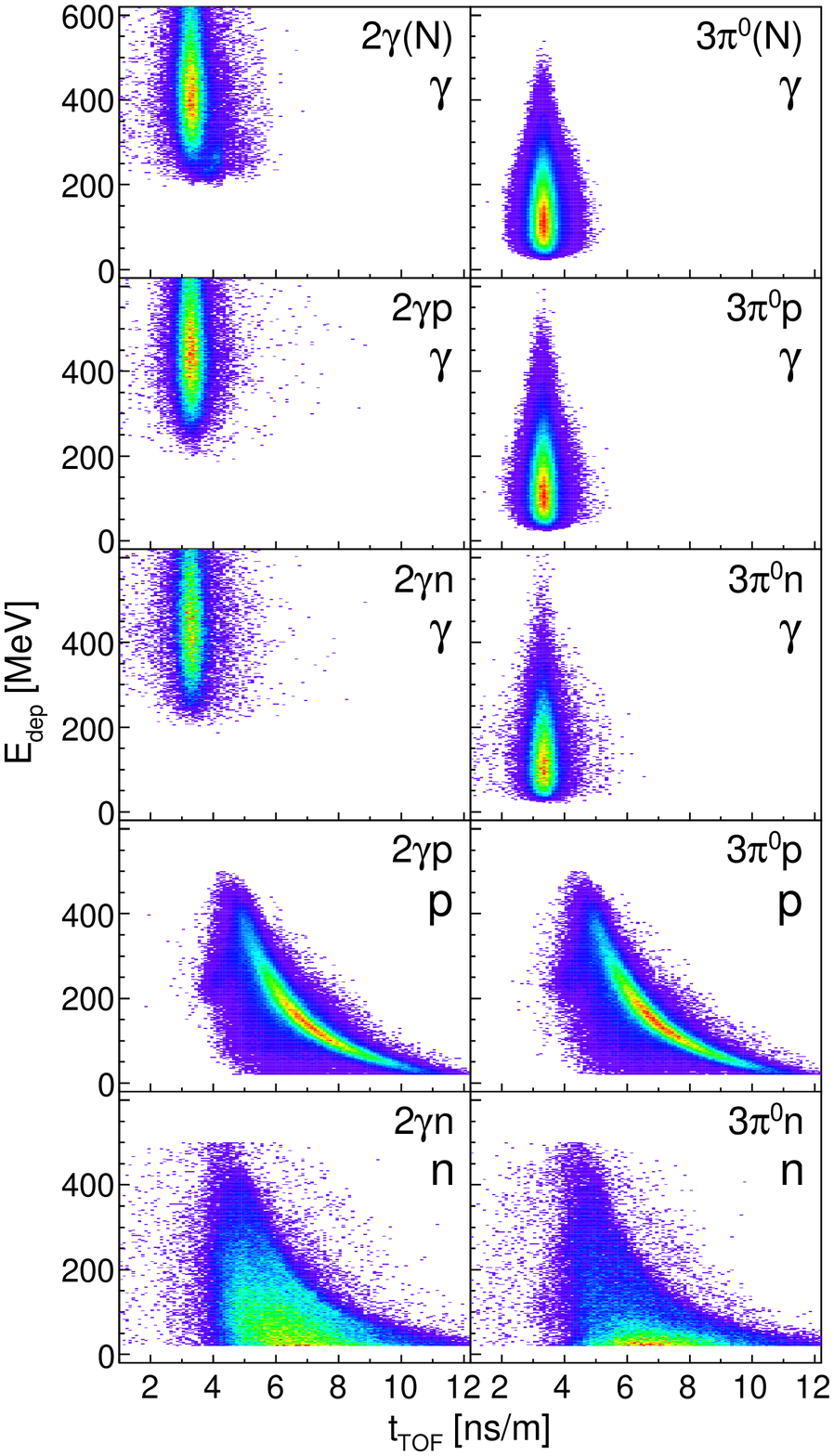}
\includegraphics[width=0.47\textwidth]{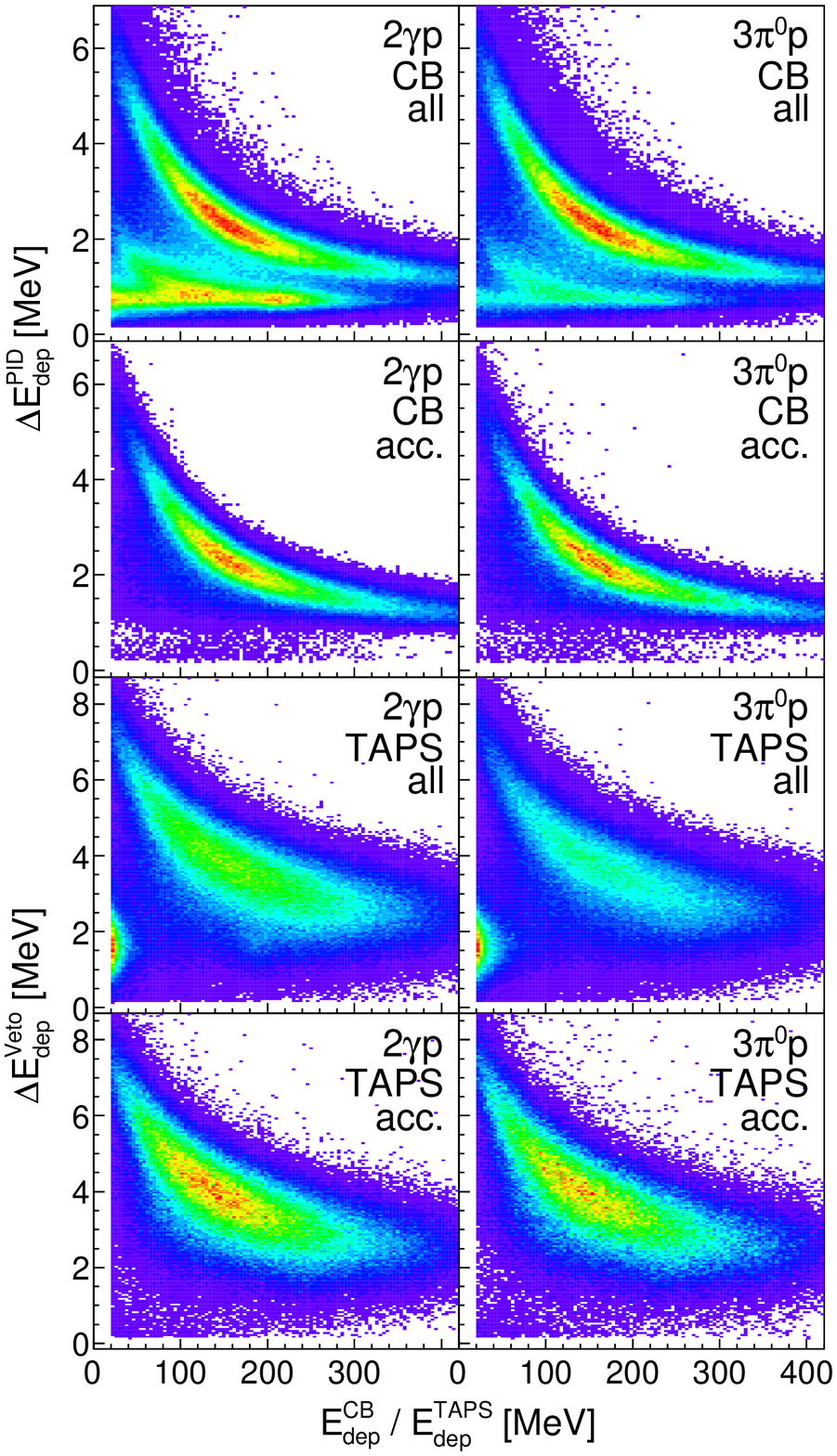}
\caption{(Color online) Left side: Time-of-flight analysis (TOF) distributions for particles detected 
in TAPS: Left column: $\EtaTwoG$ analyses. Right column: $\EtaSixG$ analyses. 
Top three rows: photon candidates in the inclusive and the two exclusive analyses. 
Bottom two rows: proton and neutron candidates in the exclusive analyses. 
Counts increase from violet to red. \\
Right side: $\Delta E$--$E$ distributions for protons: Left column: $\EtaTwoG$ analyses.
Right column: $\EtaSixG$ analyses. First row: protons in the CB (all events). 
Second row: protons in the CB (accepted events). Third row: protons in TAPS (all events). 
Fourth row: protons in TAPS (accepted events). Counts increase from violet to red.}
\label{fig:tof_deltaE}
\end{figure*}
 
So far only the information from the charged particle identification detectors 
(PID and TAPS-Veto) and the $\chi^2$ analysis have been used for particle identification.
Further constraints on particle types can be obtained for hits in TAPS from a pulse-shape 
analysis (PSA) and a time-of-flight versus energy analysis, and for hits in TAPS and in 
the CB from $\Delta E -E$ analyses and from cluster-size analyses. 

The PSA uses the fact that the pulse shape produced by protons and neutrons in BaF$_2$ differs
strongly from the signals coming from photons. This is due to the different mechanisms of 
energy deposition by these particles in matter. They result in different contributions 
to the slow ($\tau=650$ ns) and the fast ($\tau=0.9$ ns) scintillation light components 
of BaF$_2$. Therefore, in the TAPS data acquisition the signals are integrated over two ranges 
(short gate: 40 ns, long gate: 2 $\mu$s) giving two signal integrals---one containing 
mostly the short component and one containing the total signal. After an energy calibration 
based on photon signals, the short-gate energy $E_s$ and the long-gate energy $E_l$ can be 
compared to separate different particles. Convenient distributions are obtained by plotting 
the PSA-radius $r_{\mathrm{PSA}}$ versus the PSA-angle $\phi_{\mathrm{PSA}}$ using the 
transformations
\begin{equation}
r_{\mathrm{PSA}} = \sqrt{E_s^2 + E_l^2} \quad \mathrm{and} \quad \phi_{\mathrm{PSA}} = \arctan (E_s / E_l).
\end{equation}
The distributions for photon and nucleon candidates for all analyses are shown in 
Fig.~\ref{fig:psa}. Since the calibration of $E_s$ was made by setting $E_s = E_l$ for 
photons, the photons are located at $\phi_{\mathrm{PSA}} = 45^\circ$ for all PSA-radii. 
The different mean energies of the decay photons originating from the $\EtaTwoG$ and the 
$\EtaSixG$ decays are clearly represented by larger and smaller PSA-radii, respectively. 
Protons and neutrons are located at lower angles in bands showing a characteristic energy 
dependence. All analysis cuts discussed above were applied to the plotted distributions and 
very little background contamination is visible, hence the influence of the applied PSA 
cuts on the event selection is rather small. Some noticeable contamination was found in 
the nucleon spectra (in their lower, right corners) where low energy photons or electrons, 
which did not activate the veto detectors, are visible. However, no significant residue
of the photon band was observed in these spectra.

With respect to the photon mean positions $c_\gamma(r_{\mathrm{PSA}}) \approx 45^\circ$, 
cuts were established by fitting $\phi_{\mathrm{PSA}}$-distributions for different bins of 
PSA-radii. Photons were then only accepted within a $r_{\mathrm{PSA}}$-dependent $3\sigma$ 
band around $c_\gamma(r_{\mathrm{PSA}})$. Accepted nucleons had to be located at smaller 
angles than the left photon cut position for 
$r_{\mathrm{PSA}} < 85$ MeV and $r_{\mathrm{PSA}} > 380$ MeV. 
For PSA-radii between these two values no cut was applied because high energy 
punch-through nucleons were located in this area. The cuts were kept so conservative because
the background level already established by the other cuts was low and because
the PSA analysis could not be included in the MC simulations as 
modeling of the two light components of BaF$_2$ is not available.

Additional information on the detected particles provided by the various detectors, 
although not used for the application of cuts, was checked for signs of any deficiencies 
in the event selection. 

Because of the fast response of BaF$_2$ the distance from the target was sufficient for TAPS
to provide a useful time-of-flight (TOF) measurement.
The deposited energy plotted versus inverse velocity, $t_{\mathrm{TOF}}$ [ns/m], shows 
distinct distributions for the different particle types (left side of Fig.~\ref{fig:tof_deltaE}). 
Photons are located around 3.3 ns/m and the different energy of photons from the $2\gamma$ and 
$3\pi^0$ decays is again clearly visible. For protons a fairly tight correlation between inverse 
velocity and deposited energy can be seen.
Neutrons are located above 4 ns/m and do not show any correlation between 
time-of-flight and deposited energy (because the latter is more or less random).
The neutron spectra at the bottom of Fig.~\ref{fig:tof_deltaE} show no residual trace of the
proton band indicating good separation between protons and neutrons in TAPS. 
Actually, none of the spectra shows significant background structures from other 
particle species, which demonstrates the good event selection by the previously discussed 
analysis cuts.

Detected proton candidates could additionally be checked using the deposited energy in 
the PID and Veto detectors. On the right side of Fig.~\ref{fig:tof_deltaE} the 
corresponding distributions for candidates in the CB and TAPS are shown for both $\eta$ 
decay channel analyses. The spectra labeled `all', where no analysis cuts were applied, 
can be compared to the spectra `acc' representing the accepted events after all cuts. 
In the case of the CB, large background contaminations from charged pions and electrons are cleanly 
removed in the analysis. Resolution in TAPS is worse due to inferior optical coupling of 
the Veto scintillators; nevertheless, signatures of protons and electrons can be clearly 
identified. The latter are also effectively suppressed by the analysis cuts.

\begin{figure}[t]
\centering
\includegraphics[width=0.48\textwidth]{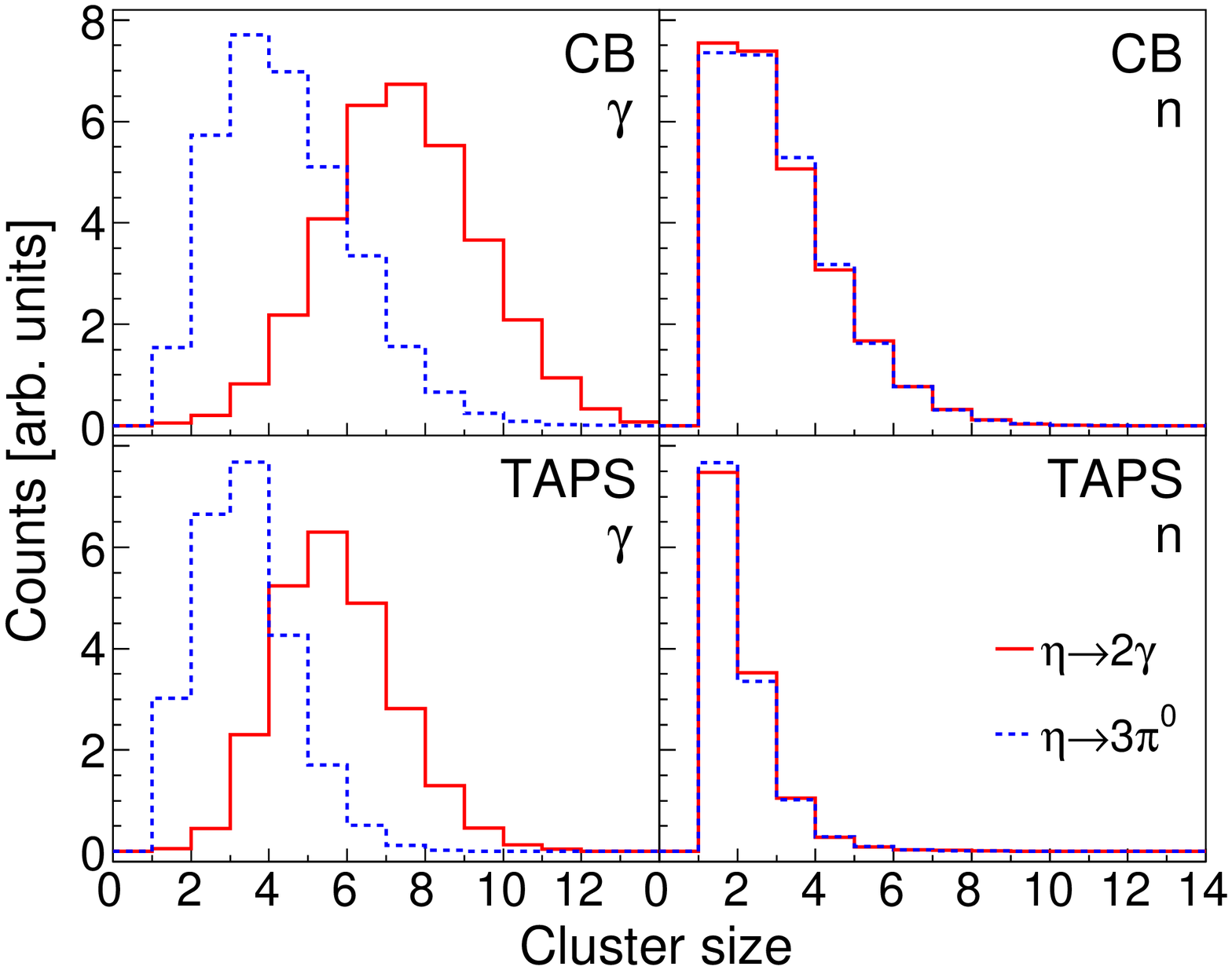}
\caption{(Color online) Cluster size distributions in the $\ReacExcNTwoG$ (solid red histograms) and the 
$\ReacExcNSixG$ (dashed blue) analyses: Top row: particles in the CB. Bottom row: particles in 
TAPS. Left column: photons. Right column: neutrons. }
\label{fig:clsize}
\end{figure}

There is no direct event-by-event discrimination of photons and neutrons in the CB
(no PSA and TOF has poor resolution due to the short flight path).
The separation is entirely based on the $\chi^2$ analysis of the invariant masses
of the `photon' pairs. However, on average there is a distinction between photon
and neutron hits by the size of the corresponding clusters, i.e., the number of
activated detector modules. Most neutron clusters consist of four or fewer detector 
elements while high energy photons (from the $\eta\rightarrow2\gamma$ decay) produce 
clusters of up to twelve crystals. The mean energy of the $\eta$ decay photons from 
the $\EtaSixG$ channel is smaller, therefore they produce smaller cluster sizes more
similar to neutrons. The measured cluster size distributions for the CB and TAPS are 
shown in Fig.~\ref{fig:clsize}. 
As expected the cluster size distributions for photons are quite different for the two
decay channels while they are very similar for neutrons.
This is a strong indication that a clean photon-neutron separation 
was achieved, even for the CB where no PSA or TOF information could be used.

\subsection{Final yield extraction}

\begin{figure*}[t]
\centering
\includegraphics[width=0.99\textwidth]{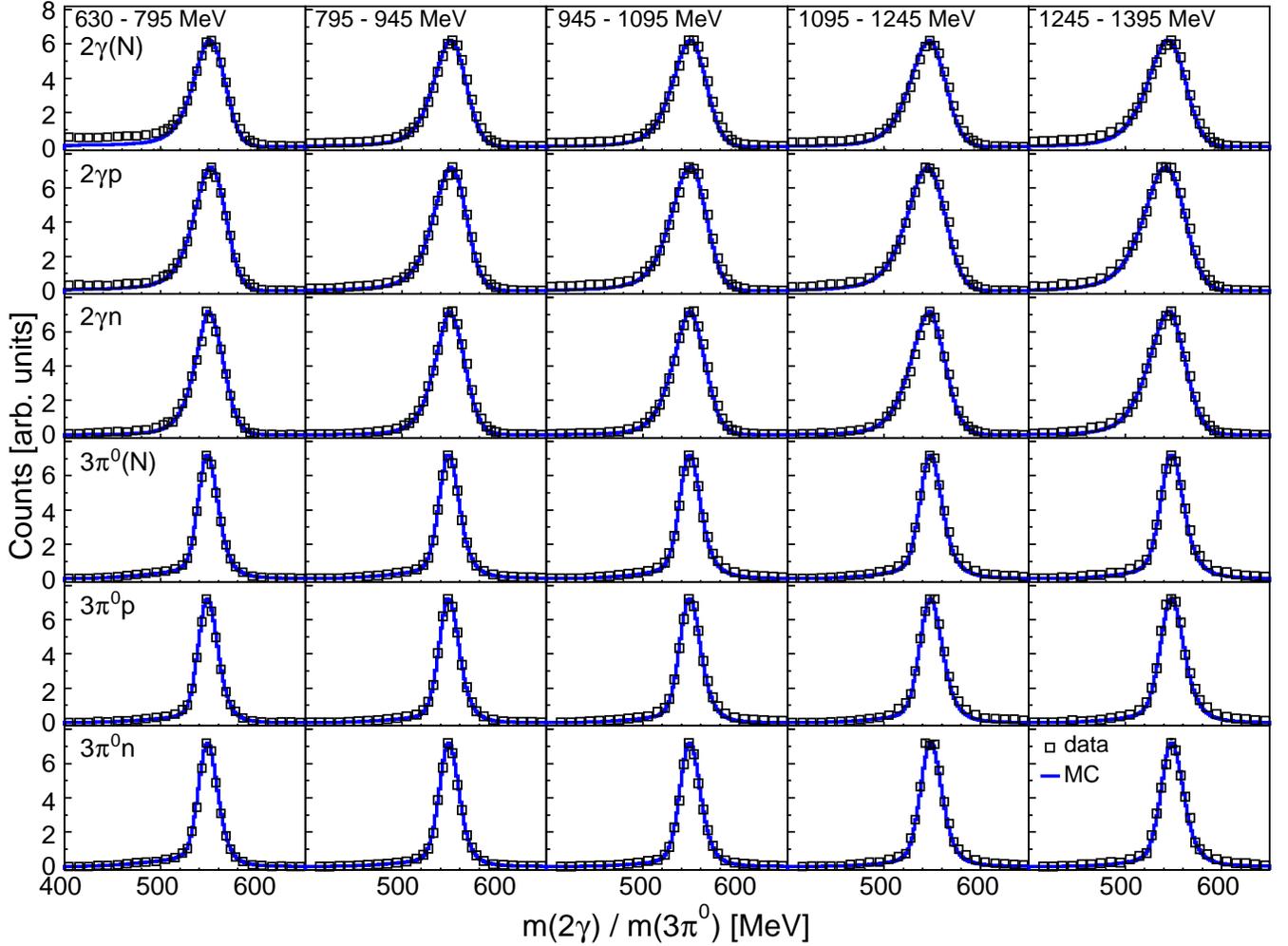}
\caption{(Color online) Invariant masses $m(2\gamma)$ and $m(3\pi^0)$ used for the identification of the 
$\eta$ mesons: Top three rows: $\EtaTwoG$ analyses. Bottom three rows: $\EtaSixG$ analyses. 
Columns: bins of incident photon energy $E_\gamma$. Black squares: experimental data. 
Blue curves: simulation.}
\label{fig:inv_mass}
\end{figure*}

The final yields were extracted from invariant-mass spectra after the application of the 
cuts discussed above. Typical examples for the different reaction types for some
energy ranges are shown in Fig.~\ref{fig:inv_mass} and compared to the distributions
obtained with MC simulations. The invariant-mass peaks from the
exclusive analyses are almost background free. Small background components 
are most visible for the $\eta\rightarrow 2\gamma$ decay in the inclusive reaction.
The residual backgrounds were subtracted by a fit made using the peak shape from the 
simulated distribution together with a polynomial background.
For the extraction of angular distributions, the entire analysis 
procedure was of course done separately for each data point of the angular distribution.

\subsection{$W$ reconstruction as final-state invariant mass}
\label{sec:w_rec}
A full reconstruction of the kinematics was performed for the exclusive analyses. 
This method allowed a calculation of the Fermi momentum of the participant nucleon in the 
initial state. A cut on the momentum rejecting events with momenta above 80 MeV was used 
in a special version of the analysis (later referred to as type II) which attempted to optimize
the resolution of the $W$ reconstruction. In the  standard analysis no such 
cut was applied, partly due to the loss of statistics and because it introduces additional
systematic effects in the extraction of the  cross section close to threshold, where 
larger Fermi momenta play a crucial role.

Quasifree cross sections calculated as functions of the photon beam energy $E_\gamma$ are 
affected by the Fermi motion of the initial-state nucleons, which are
bound inside the deuteron. This means that a fixed value of $E_\gamma$ corresponds to a 
distribution of center-of-mass energies $W$ due to the convolution with the Fermi momentum 
distribution. The resulting cross sections are then smeared compared to the fundamental 
cross sections at fixed values of $W$. This loss of resolution is mainly a problem when 
sudden changes occur in the latter, as in $\eta$ photoproduction at threshold and in 
the region of interest around $W = 1680$ MeV. Therefore, in this work the `true' center-of-mass 
energy was calculated by a full reconstruction of the reaction in impulse approximation 
from the final state \cite{Jaegle_11}. For a limited angular region, $W$ could additionally 
be reconstructed using a time-of-flight measurement of the recoil nucleons.

From the measurement of the energies and directions of the decay photons, the four-momentum 
of the $\eta$ meson could be completely reconstructed. The same is in principle possible 
for final-state protons although a special calibration would be needed to deduce the kinetic 
energy from the energy deposition in the detector. As there is no correlation between the 
kinetic energy of neutrons and their deposited energy, a reconstruction of the neutron 
energies was not possible in this way. The energy of the recoil nucleon was thus treated 
as unknown in the reconstruction with only the measured angles used to reconstruct
the direction of its momentum. This was done in the same way for protons and neutrons
in order to maintain identical systematic uncertainties. Additional unknown quantities 
are the three components of the spectator nucleon momentum in the final state, 
leaving in total four unknown quantities since $E_\gamma$, all masses of the involved 
particles, and the momentum of the incident deuteron (at rest) are known.
These four variables were determined from the four equations following from energy and
momentum conservation. In this way, the three momenta of final-state participant (and spectator)
nucleon and the final-state invariant mass of the recoil-nucleon-meson pair could be 
reconstructed in the plane-wave impulse approximation. 

The kinetic energy of nucleons detected in TAPS could also be determined via a 
time-of-flight measurement. Regarding the different time resolutions of the detectors, the 
best measurement would be possible for events with a photon in TAPS along with the recoil 
nucleon. Due to the reaction kinematics and the experimental trigger, there are practically 
no such events in the $\EtaTwoG$ analyses and only results with low statistics
could be extracted in the $\EtaSixG$ analyses. Therefore, the main TOF results were deduced from time 
measurements relative to the tagger, which are affected by an inferior time resolution. 
TOF measurements were not possible for the CB because of the short flight path and the poor time 
resolution in the NaI(Tl) crystals. Due to this restriction, $W$ could only be reconstructed
for $\eta$ polar angles $\theta_{\eta}^{*}$ in the center-of-mass frame with $\CTTOF$. 
Nevertheless, this independent method serves as a check for the $W$ reconstruction
discussed above. 
 
\begin{figure}[t]
\centering
\includegraphics[width=0.48\textwidth]{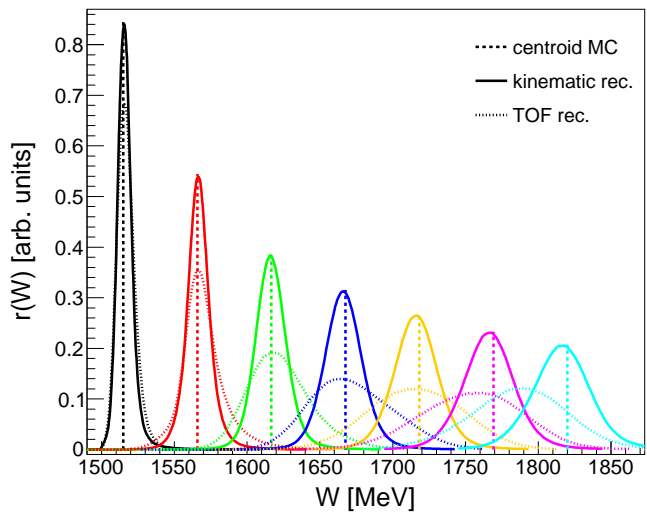}
\caption{(Color online) Resolution of the $W$ reconstructions illustrated by the response curves
of the kinematic reconstruction (solid curves) and the 
reconstruction via time-of-flight (dotted curves) at several discrete values of $W$ 
(dashed lines). The response curves were obtained from the $\ReacExcNTwoG$ simulation.}
\label{fig:wres}
\end{figure}

Cross sections as a function of $W$ reconstructed with the two discussed methods are no longer
affected by nuclear Fermi motion but depend on the experimental resolution for the reconstructed
$W$. The resolution was investigated with MC simulations of the experimental setup (see 
section \ref{sec:cs_extr}). Phase-space events at several fixed values of $W$ 
($\delta$-functions) were generated and the responses of the detectors were modeled. 
The same analysis as used for experimental data was applied to the simulated data and
the resulting distributions for $W$ from the $\ReacExcNTwoG$ analysis are summarized in 
Fig.~\ref{fig:wres}. 

The effects from the energy and angular resolution for the 
$\eta$ mesons enter into both analyses in the same way for the determination of
the $\eta$ four-momentum. The resolution for the angle of the recoil nucleon also enters
into both analyses via the definition of the direction of the nucleon momentum.
All these factors increase the width of the observed distributions with rising $W$.
The angular resolution for the recoil nucleon degrades significantly above $W \sim 1550$ MeV
where the majority of nucleons are detected in the CB for which the polar angle resolution 
$\Delta\theta$ is worse than in TAPS ($\Delta\phi$ being similar).
The final ingredient, the kinetic energy of the recoil nucleon, is determined from 
the kinematic reconstruction or from the TOF measurement.  It is evident from 
Fig.~\ref{fig:wres} that the first method results in better resolution, in particular for
larger values of $W$. For the TOF reconstruction for higher energies even the centroids 
are shifted. This is due to the fact that the TOF resolution for neutrons is not very good
(additional uncertainty is introduced because they can interact at any depth in the
crystals and the TOF flight path is not very long) and that, at larger kinetic energies,
the TOF-energy relation becomes so flat that small effects in TOF result in large uncertainties
for the energies.  

In the case of the kinematic reconstruction, the corresponding 
resolution for the $\EtaSixG$ analysis is slightly better---as it also is for the proton
analyses. When approximated with Gaussians they all show a nearly linear rise from 
$\Delta W(\mathrm{FWHM}) \sim 10$ MeV at 1515 MeV to $\Delta W \sim 40$ MeV at 1820 MeV. 

In summary, a FWHM resolution of 30 MeV in the region of interest around $W = 1680$ MeV 
could be achieved with the kinematic reconstruction of the final-state invariant mass,
while the TOF reconstruction yields an inferior resolution of about 70 MeV.

\subsection{Extraction of cross sections}
\label{sec:cs_extr}
The obtained yields were normalized to differential cross sections by using the target 
surface density of 0.2304 nuclei/barn (0.147 nuclei/barn for one of the beam times), the 
flux of the incoming photon beam, the analysis-dependent detection efficiency and the 
$\eta$ decay branching ratios $\Gamma_{2\gamma} = 39.41\%$ and $\Gamma_{3\pi^0} = 32.68\%$, 
respectively \cite{PDG_12}.

\begin{figure}[b]
\centering
\includegraphics[width=0.48\textwidth]{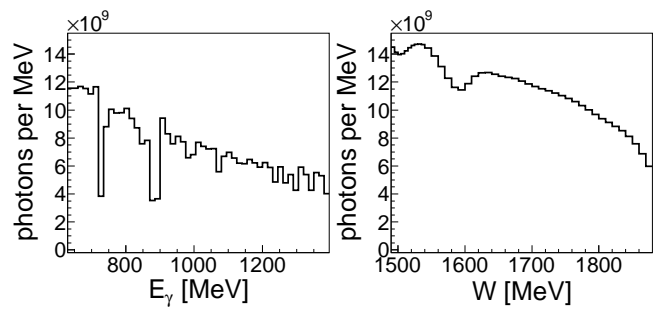}
\caption{Photon flux on the target for the May 2009 dataset: 
Left side: measured photon flux as a function of $E_\gamma$. \\
Right side: flux as a function of $W$ obtained from folding with the deuteron Fermi 
momentum distribution.}
\label{fig:flux}
\end{figure}

\begin{figure*}[t]
\centering
\includegraphics[width=0.99\textwidth]{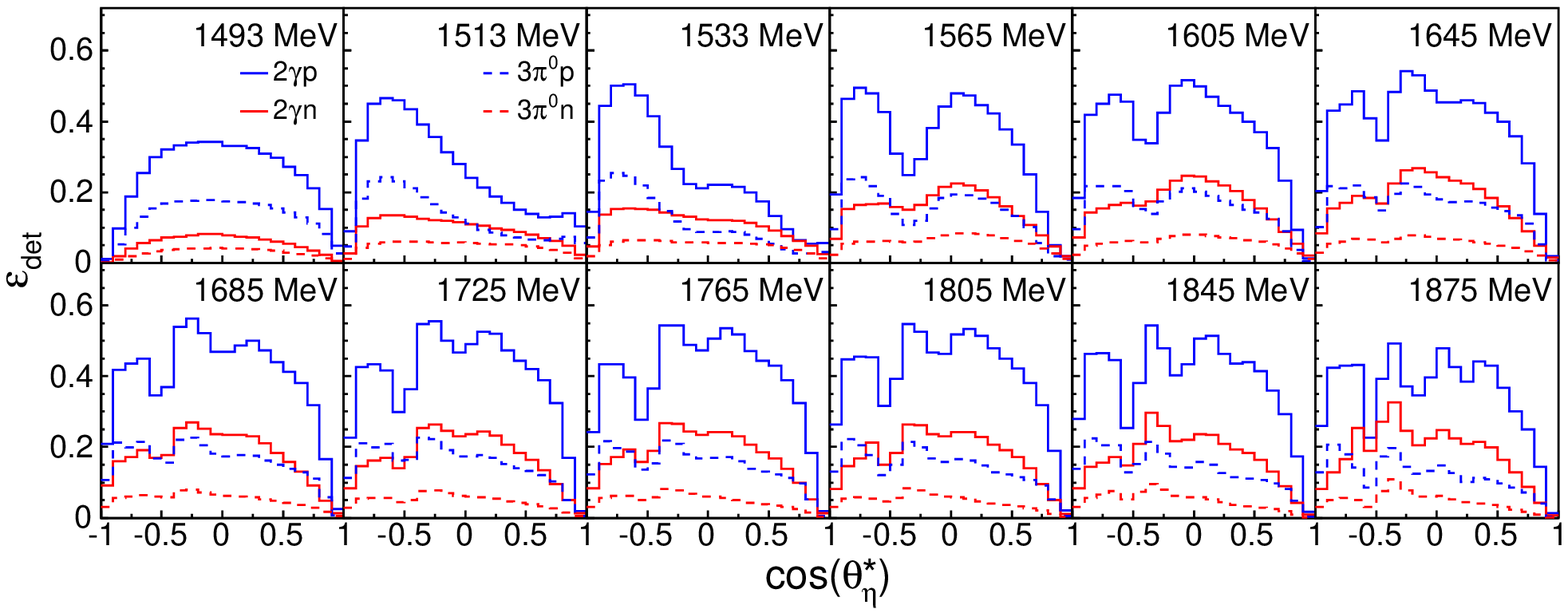}
\caption{(Color online) Detection efficiencies as a function of $\CT$ for the exclusive analyses for 
different bins of $\WK$: Solid curves: $\EtaTwoG$ proton (blue) and neutron (red) analyses. 
Dashed curves: $\EtaSixG$ proton (blue) and neutron (red) analyses.}
\label{fig:deteff}
\end{figure*}

Yields for all bins in $(E_\gamma,\CT)$, $(\WK,\CT)$ and $(\WT,\CT)$ (with $\CT$ evaluated 
in the corresponding center-of-mass frame) were individually extracted by integrating the 
appropriate $2\gamma$ or $3\pi^0$ invariant-mass histograms. Having applied all analysis cuts 
as discussed in Sec.~\ref{sec:iden}, these histograms were background-free for the 
exclusive analyses. In the case of the inclusive analyses, the background was non-negligible 
since no cuts on the recoil nucleon could be applied. Here, the signals were extracted 
using a fit of the distributions consisting of the combined peak shape from the simulated distribution 
and a second-order polynomial function for the background. Energy (but not $\CT$) dependent 
contributions to the yields originating from the target windows were subtracted using data 
that were measured while the target cell was empty. These contributions were about 
5--7\% and showed a rather smooth energy dependence.

The photon flux on the target as a function of $E_\gamma$ was calculated via 
\begin{equation}
N_{\gamma}(E_{\gamma}) = N_{e^-}(E_{\gamma}) \epsilon_{tg}(E_{\gamma}).
\end{equation}
The number of electrons $N_{e^-}(E_{\gamma})$ in the photon tagger was counted during the 
whole experiment. The so-called tagging efficiency $\epsilon_{tg}(E_{\gamma})$, i.e.,
the fraction of correlated photons passing through the beam collimator, was determined in
frequent, dedicated measurements at low beam intensity. Running at these conditions
ensured that random electron coincidences were minimized, and that the photon detection 
efficiency of the lead-glass detector, which was moved into the photon beam, was still 
close to 100\%. Besides these absolute values of the tagging efficiency, relative values
were available at all times from the measured relative beam intensity using an
ionization chamber placed at the end of the photon beam line. By normalization of
the relative values to the absolute measurements, a time-dependent tagging efficiency
was calculated. The resulting flux 
integrated over one of the beam times is shown as a function of $E_\gamma$ at
the left side of Fig.~\ref{fig:flux}. For the normalization of the cross section 
obtained using either the kinematic or the TOF reconstruction, an effective photon flux as 
a function of $W$ had to be calculated. The effective distribution of $W$ values was 
calculated by folding the incoming photon beam energy distribution with the nucleon 
momentum distribution inside the deuteron. For the latter the wave function of the Paris 
$N$--$N$ potential was used \cite{Lacombe_81}. The resulting flux is shown at the right
side of Fig.~\ref{fig:flux}.

The detection efficiency was determined with a Geant4-based model \cite{Agostinelli_03}
of the experimental setup. Events covering the complete phase space of quasifree $\eta$ 
production were generated and tracked by the simulation. The resulting detector information 
was analyzed using the same analysis as for real data. In addition, the experimental 
trigger conditions had to be modeled realistically. In fact, an even more restrictive 
implementation of the CB energy sum trigger was imposed (also on the experimental data), 
namely that only the decay photons were allowed to contribute. The same restriction was 
implemented for the hit multiplicity to avoid systematic differences in the proton and 
neutron analyses due to the different interaction of these particles with the detectors.

\begin{figure*}[t]
\centering
\includegraphics[width=0.95\textwidth]{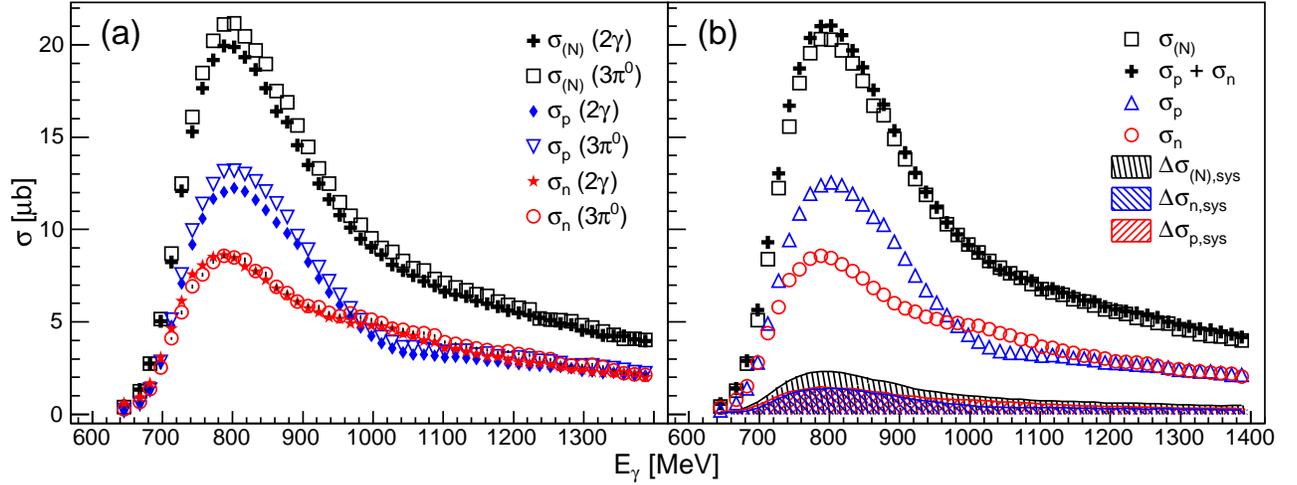}
\caption{(Color online) Total cross sections as a function of the photon beam energy $E_\gamma$: 
(a) Comparison of the two $\eta$ decay analyses of the proton 
(blue diamonds and triangles), neutron (red stars and circles) and the inclusive analyses 
(black crosses and squares).
(b) Comparison of the averaged proton (blue triangles) and neutron (red circles) results
and of their sum (black crosses) to the inclusive data (black squares). 
Hatched areas: total systematic uncertainties of inclusive (black), proton (blue) and 
neutron (red) data.}
\label{fig:tcs}
\end{figure*}

Subsequently, the detection efficiency was calculated as the ratio of detected and 
generated events for each bin of the excitation functions. Some examples are shown in 
Fig.~\ref{fig:deteff} for the exclusive analyses. Special attention was given to the 
detection efficiencies of the recoil nucleons, as the systematic uncertainties of the hadronic 
models in the energy range covered by this experiment were suspected to be rather large. 
Especially the tracking of low energy neutrons through different materials requires 
specific and accurate cross sections for the nuclear reactions involved. The proton 
efficiencies are highly sensitive to the modeled detector geometries and material budgets 
as well. A measurement on a hydrogen target was used to check and correct the 
nucleon detection efficiencies obtained by simulation. For this purpose, relative 
corrections of the simulated nucleon efficiencies for the free reactions $\ReacExcP$ and 
$\ReacExcNPizPip$ were deduced from a comparison of hydrogen experimental data and the 
corresponding simulation, and applied in the deuteron analyses. For nucleons detected in 
TAPS corrections for recoil protons (neutrons) were on average around +7.3\% (+12.1\%)
(the detection efficiency was overestimated by the simulation), while for the CB
they were -1\% (-3.5\%) (underestimated on average by simulation). In the gap region between 
the CB and TAPS the corrected nucleon efficiencies were found to be still inaccurate. 
This corresponds to values of $\CT$ around $-0.6$ for $W > 1550$ MeV where the 
sharp efficiency dependence on $\CT$, especially for the proton, 
can be clearly seen in Fig.~\ref{fig:deteff}. 
As a solution, differential cross sections as functions of the lab polar angle of the 
nucleons were interpolated in the problematic regions and correction factors were calculated
that were finally applied on an event-by-event basis in the analysis. 

\subsection{Systematic uncertainties}
Common to all results are the global systematic uncertainties of the photon flux (3\%), 
the target surface density (4\%), the $\eta$ decay branching ratios ($<$ 1\%) and the 
approximately constant uncertainty of the empty target subtraction (2.5\%). The systematic 
uncertainty in the photon flux mainly comes from the absolute measurements of the tagging 
efficiency and was estimated by the extreme values of the normalization of the relative 
flux measurements to the absolute measurements. The target surface density depends on the length 
of the target cell, which is subject to deformations when the target is cooled 
down. The systematic uncertainties of the $\eta$ decay branching ratios are almost 
negligible \cite{PDG_12}. Due to the low statistics of the empty target runs, 
a conservative estimate of roughly half the relative yield contribution (2.5\%) was made.

Several systematic uncertainties were found to be of rather different importance for the 
various analyses and showed a strong energy and $\CT$ dependence. Therefore, they were 
calculated individually and for all bins of the obtained cross sections. First, the CB 
energy sum trigger was found to be of great importance especially for the $\EtaSixG$ 
analysis. Its uncertainty was estimated by slightly different applications of the software 
model trigger in the analysis of simulated data. All analysis cuts were varied by $\pm3\%$ 
and systematic uncertainties were estimated from the differences between the results. 
Uncertainties in the nucleon detection efficiencies were estimated taking into account effects of 
different hadronic models in Geant4, trigger and cut effects in the analysis of the 
hydrogen data used for the efficiency correction, and the influence of the correction applied for
the data corresponding to the TAPS-CB gap region.

\begin{figure*}[p]
\centering
\includegraphics[width=0.9\textwidth]{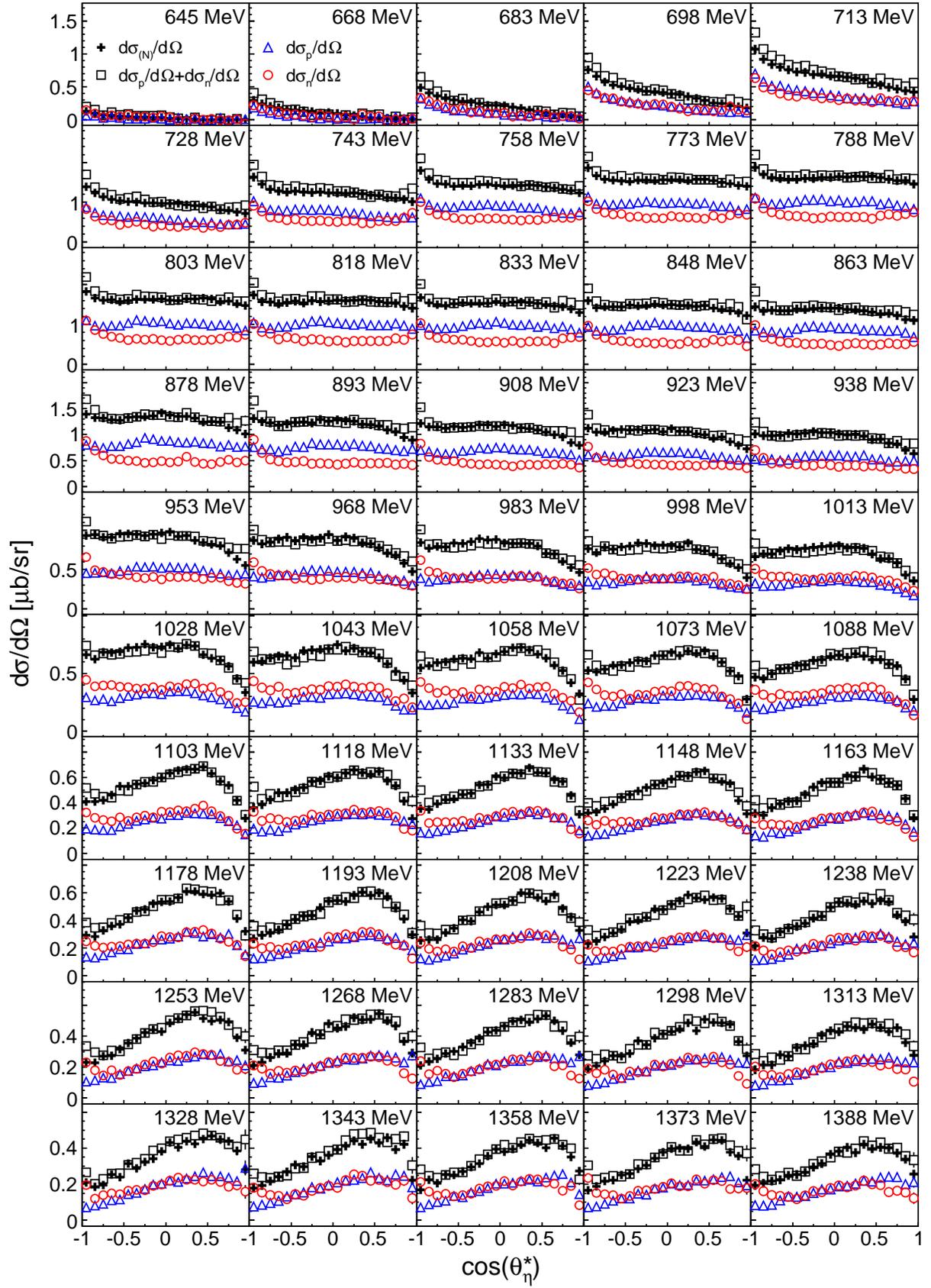}
\caption{(Color online) Differential cross sections as a function of the photon beam energy $E_\gamma$ 
and $\CT$: Blue triangles: exclusive proton. Red circles: exclusive neutron. 
Black crosses: sum of proton and neutron data. Black squares: inclusive data.}
\label{fig:dcs}
\end{figure*}

The many small systematic uncertainties from the different analysis steps were added quadratically
and the result was added linearly to the uncertainties of the photon flux and the target
density.
The total uncertainty for the inclusive $\EtaTwoG$ analysis shows almost no energy and 
$\CT$ dependence and is about 7\%. The total uncertainty for the corresponding 
$\EtaSixG$ analysis falls from 11\% at threshold to 7\% at $E_\gamma = 1$ GeV above which
it is constant. The reason for these increased values are the higher uncertainties in the 
backward region of $\CT$, which are related to the CB energy sum trigger. The systematic 
uncertainties for the proton analyses are almost energy and $\CT$ independent (with the exception 
of the most forward angular bin and the CB-TAPS gap region located at $\CT\sim-0.5$ for higher values of $W$) 
and are about 6--7\%. The systematic uncertainties 
for the neutron analyses show a more pronounced energy and $\CT$ dependence. The total values 
for the $\EtaTwoG$ ($\EtaSixG$) analysis are around 12\% (15\%) at threshold, have a local 
maximum of 13\% (14\%) near $W = 1580$ MeV, and fall more or less linearly to 9\% (10\%) 
at the maximum energy.

\section{Results}
The results presented in this section were obtained by combining the datasets from all
three beam times. Furthermore, the data of the $\EtaTwoG$ and $\EtaSixG$ 
analyses were averaged according to their statistical weights to calculate the final cross 
sections. Differential cross sections were extracted as functions of $(E_\gamma,\CT)$, 
$(\WK,\CT)$ and $(\WT,\CT)$ where $\CT$ was always evaluated in the corresponding 
center-of-mass frame. Total cross sections were obtained by fitting the angular 
distributions with Legendre polynomials.

\subsection{Cross sections as a function of $E_\gamma$}
The total cross-section results are shown in Fig.~\ref{fig:tcs}. On the left side, 
the data from the two $\eta$ decay analyses are compared to each other. With the exception 
of the neutron data in the threshold region, the cross sections extracted from the $\EtaSixG$ 
analyses are slightly larger. This could be due to residual background from direct $3\pi^0$ 
photoproduction (the invariant mass spectrum of direct $3\pi^0$ production peaks in the 
S$_{11}$(1535) region close to the $\eta$-mass due to trivial kinematic relations). 
Other effects at lower 
photon beam energies could be caused by the CB energy sum trigger, whereas at higher 
energies cluster overlaps in the $\EtaSixG$ analyses could lead to systematic effects. 

At the right side of Fig.~\ref{fig:tcs}, the data averaged over the $\eta$ decays 
are shown. The inclusive result and the sum of the proton and neutron cross sections
are compared. Since the coherent production of $\eta$ mesons off the deuteron is very 
small \cite{Weiss_01}, the two exclusive 
cross sections should add up to the inclusive data. Within a range of 10\%, which is 
compatible with all of the involved systematic uncertainties, this is indeed the case. 
The good agreement is also clearly visible in the corresponding angular distributions 
that are shown in Fig.~\ref{fig:dcs}. In the region of the \SOneOne resonance 
($E_\gamma = $ 758--923 MeV) for example, 
the proton and neutron distributions are curved in opposite directions due to an 
interference with the \DOneThree with different signs \cite{Weiss_03}. Their sum is flat 
which is reproduced by the direct inclusive measurement. Also at higher energies, the angular 
distributions of sum and direct measurement agree very well. This is a strong indication 
that systematic uncertainties, although quite large in case of the neutron data, are 
generally well under control. 

\begin{figure}[t]
\centering
\includegraphics[width=0.48\textwidth]{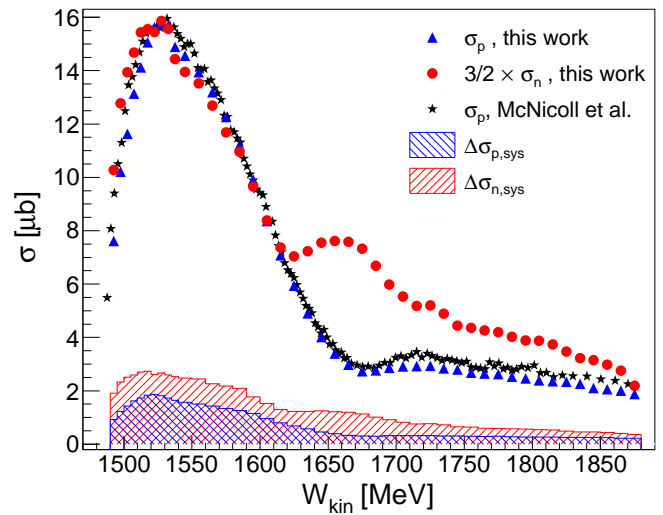}
\caption{(Color online) Total cross sections as a function of the final-state invariant mass 
$\WK=m(\eta N)$: Blue triangles: proton data. Red circles: neutron data scaled by $3/2$. 
Black stars: free proton data from MAMI-C \cite{McNicoll_10}.
Hatched areas: total systematic uncertainties of proton (blue) and neutron (red) data.}
\label{fig:tcsw}
\end{figure}

\begin{figure*}[p]
\centering
\includegraphics[width=0.99\textwidth]{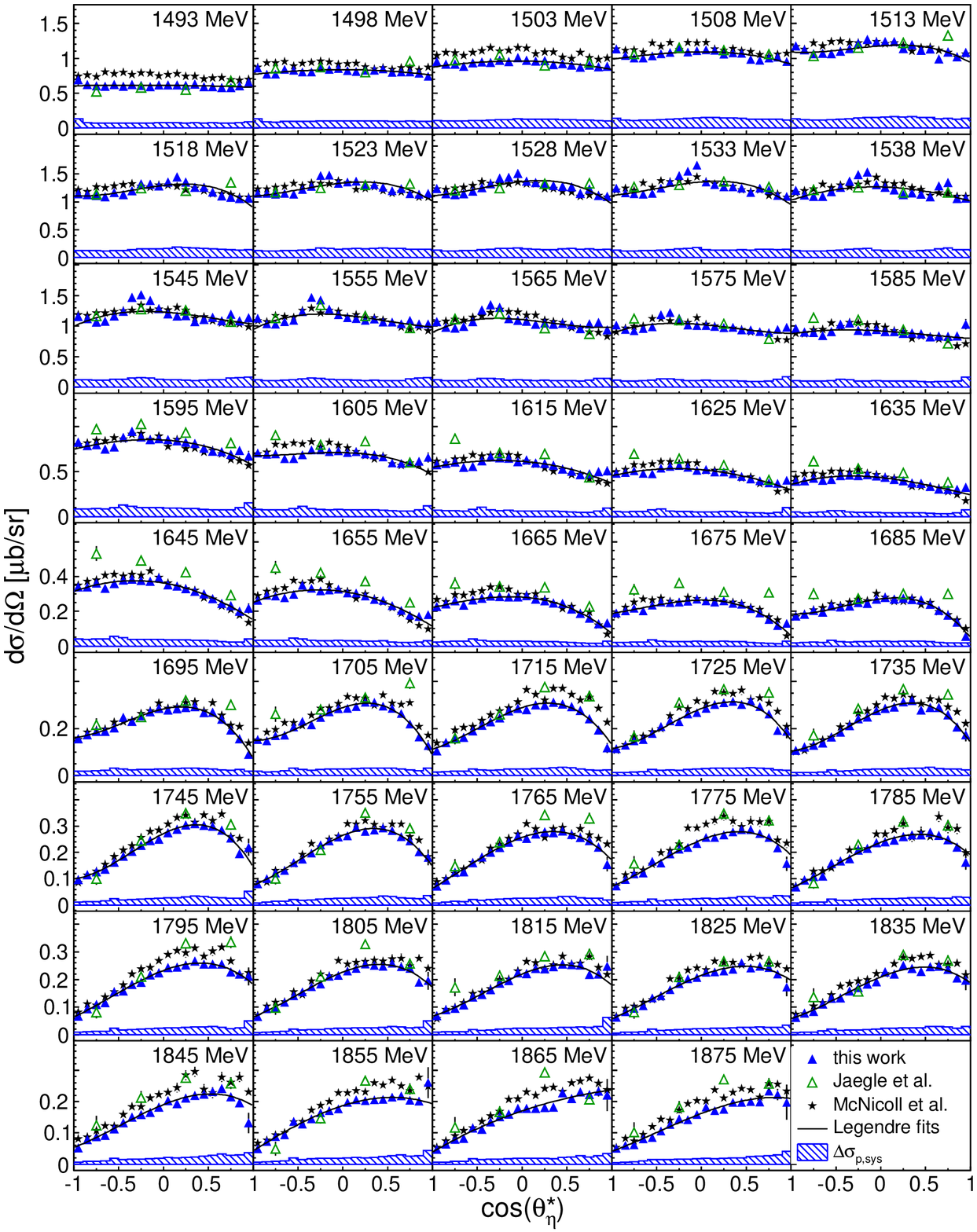}
\caption{(Color online) Differential cross sections for $\ReacExcP$ as a function of the final state invariant mass 
$\WK=m(\eta p)$ and $\CT$: Filled blue triangles: exclusive proton. 
Open green triangles: quasifree data from CBELSA/TAPS \cite{Jaegle_11}. 
Black stars: free proton data from MAMI-C \cite{McNicoll_10}. 
Black curves: Legendre fits to the present results. Hatched blue areas: total systematic uncertainties
in the present work.}
\label{fig:dcsw_p}
\end{figure*}

\begin{figure*}[p]
\centering
\includegraphics[width=0.99\textwidth]{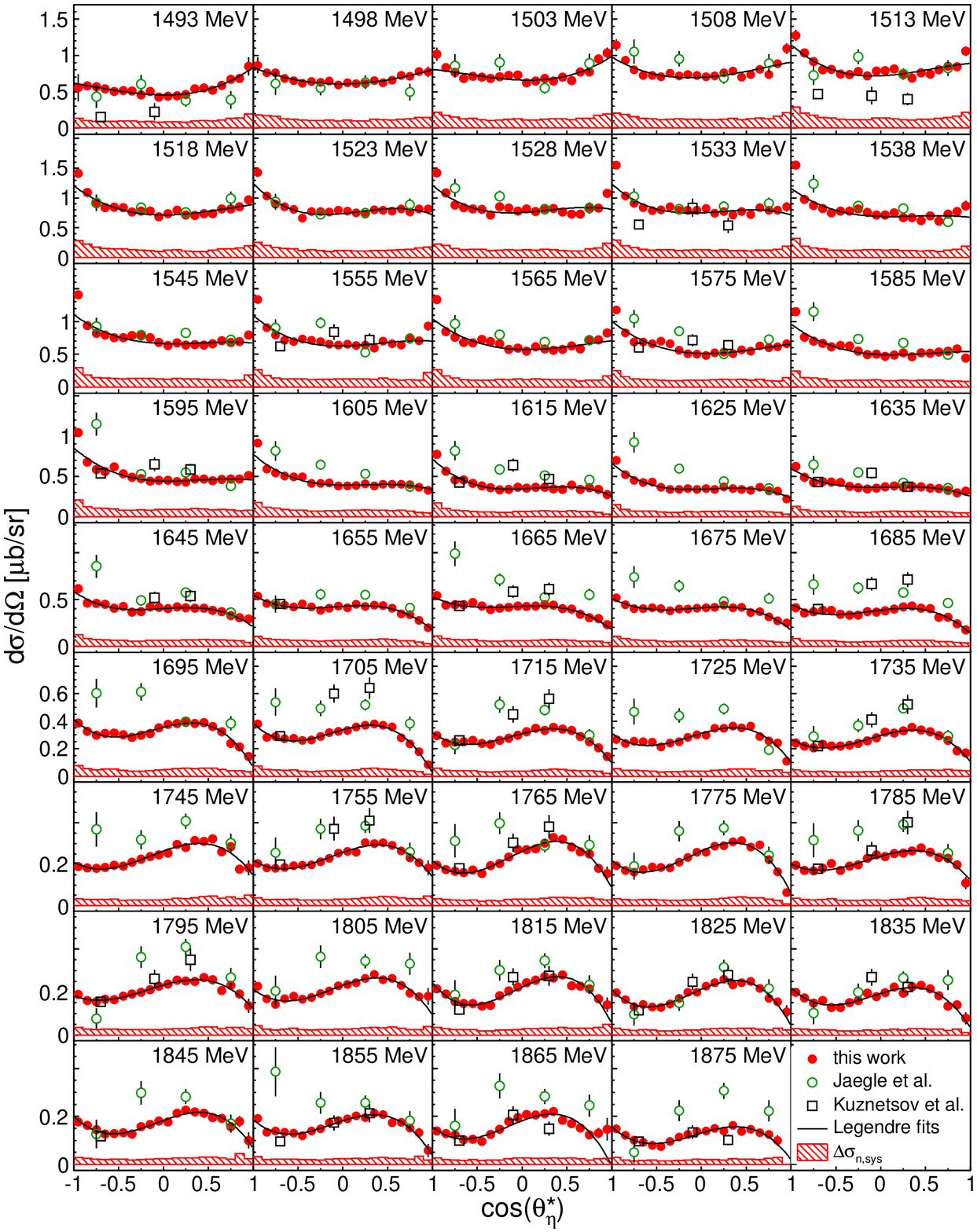}
\caption{(Color online) Differential cross sections for $\ReacExcN$ as a function of the final-state invariant mass 
$\WK=m(\eta n)$ and $\CT$: Filled red circles: exclusive neutron. 
Open green circles: quasifree data from CBELSA/TAPS \cite{Jaegle_11}. 
Open black squares: quasifree data from GRAAL \cite{Kuznetsov_07}.
Black curves: Legendre fits to the present results. Hatched red areas: total systematic uncertainties
in the present work.}
\label{fig:dcsw_n}
\end{figure*}

\begin{figure*}[t]
\centering
\includegraphics[width=0.99\textwidth]{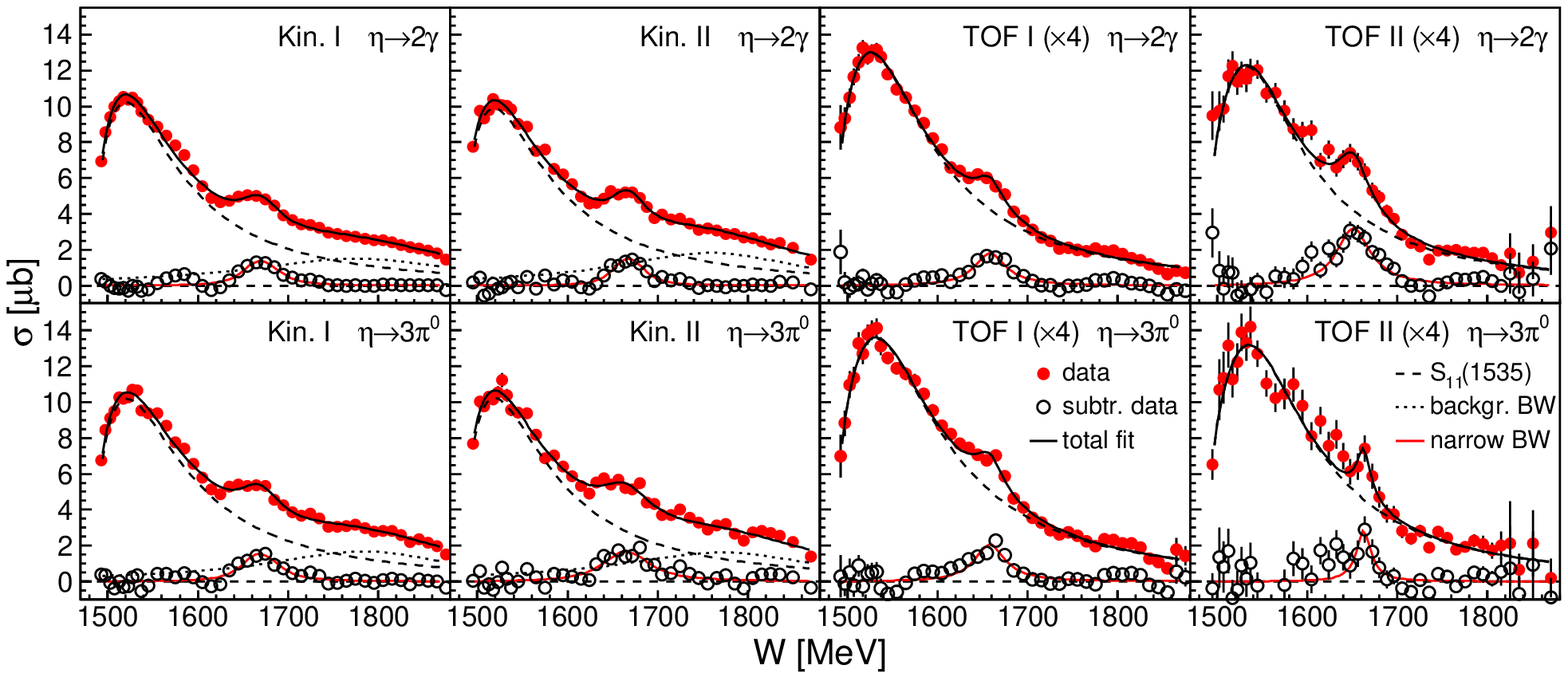}
\caption{(Color online) Phenomenological fits of the total $\ReacExcN$ cross sections as functions of $\WK$ and 
$\WT$: Top row: $\EtaTwoG$ analyses. Bottom row: $\EtaSixG$ analyses. 
Columns from left to right: Kinematic $W$ reconstruction with standard (Kin.~I) 
and more strict (Kin.~II) cuts, time-of-flight $W$ reconstruction with standard (TOF I) 
and more strict (TOF II) cuts. Points: Original data (filled red circles) and background 
subtracted data (open black circles). The TOF data were scaled by a factor of 4. 
Curves: Total fit (solid black), \SOneOne contribution (dashed black), integrated background
Breit-Wigner (dotted black) and narrow BW (solid red).}
\label{fig:pos_width}
\end{figure*}

\subsection{Cross sections as a function of $\WK$}
Fig.~\ref{fig:tcsw} shows the total cross sections as a function of the final-state 
invariant mass $\WK$ obtained using the kinematic reconstruction of the nucleon energies. 
As discussed before, no effects from Fermi motion should be present in these data---it should 
only be affected by the resolution of the $W$ reconstruction. Therefore, the proton data can be 
directly compared to data measured on the free proton target as, for example, obtained at MAMI-C 
\cite{McNicoll_10}. The main characteristic features of the latter data are reproduced.
There are some discrepancies at threshold, which are most probably due to 
the much poorer (ca.~one order of magnitude) resolution in $W$ of the current analysis and 
the complicated proton detection efficiency in the region of the \SOneOne resonance. 
Above $W = 1600$ MeV the two data sets deviate by up to 15\% for the highest measured
invariant masses. The differential cross sections for $\ReacExcP$ are shown in 
Fig.~\ref{fig:dcsw_p} and also compared to the free proton data. 
In addition, the quasifree data obtained by the CBELSA/TAPS collaboration \cite{Jaegle_11} 
are plotted. The very precise angular distributions of \cite{McNicoll_10} are in general 
well reproduced by this work. Some residual effects from the uncertain proton detection 
efficiency in the CB-TAPS gap region are still visible. They are located in the energy 
bins for 1518 MeV $< W <$ 1655 MeV, first around $\CT\approx 0.3$ and then slowly moving to
backward angles up to $\CT\approx -0.65$. The issues in the determination of these data 
points are accounted for by increased systematic uncertainties. Altogether, the quasifree and free
proton data agree quite well (for most kinematics within systematic uncertainties), which indicates 
that for this reaction channel nuclear effects from FSI are not important, so that the quasifree
neutron data can be regarded as close approximation of the free $\ReacExcN$ 
cross sections. This is by no means trivial. In a similar investigation of photoproduction of 
$\pi^0$ mesons off nucleons bound in the deuteron \cite{Dieterle_14} substantial effects (on the
order of 25\%) were found and also $\eta$ production off nucleons bound in $^3$He nuclei
is strongly affected by FSI \cite{Witthauer_13}.   

The total cross section for $\ReacExcN$ shown in Fig.~\ref{fig:tcsw} was scaled 
by $3/2$ to compensate for the known ratio $\sigma_n/\sigma_p \approx 2/3$ 
\cite{Krusche_95_2,Hoffmann_Rothe_97, Hejny_99} in the maximum of the \SOneOne resonance. 
The shapes of the corresponding peaks in the proton and neutron cross-section data are very similar.
The small deviations are caused by the different systematic 
effects in the proton and neutron detection efficiencies. Above $W = 1615$ MeV 
the neutron cross section deviates strongly from the proton results and exhibits
a pronounced peak-like structure around $W = 1670$ MeV. This structure, already observed 
by earlier measurements \cite{Kuznetsov_07,Miyahara_07,Jaegle_11}, is thus confirmed 
by this work with much superior statistical significance. 

The corresponding differential cross sections for $\ReacExcN$ are shown in Fig.~\ref{fig:dcsw_n}. 
The data from GRAAL \cite{Kuznetsov_07} and CBELSA/TAPS \cite{Jaegle_11} are plotted for 
comparison. There is reasonable agreement between all data in the region of the \SOneOne 
resonance. In the vicinity of the peak-like structure, some deviations between the different 
measurements are visible. Above around $W = 1800$ MeV, the agreement with the GRAAL data  
improves again. The much better statistical quality of the present data compared to the previous 
measurements is obvious from the figure. 

\subsection{Properties of the structure}

\begin{figure*}[t]
\centering
\includegraphics[width=0.99\textwidth]{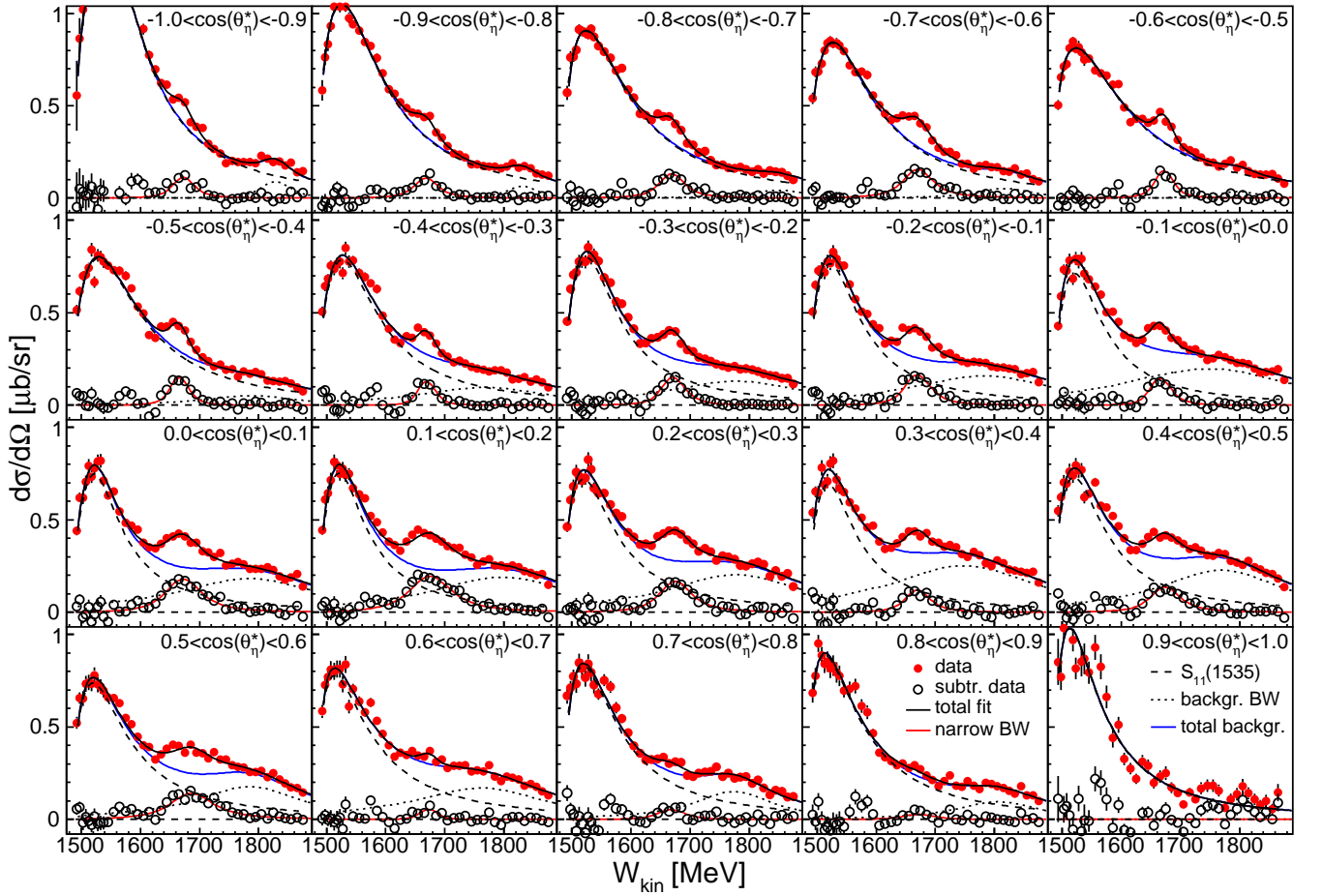}
\caption{(Color online) Differential cross sections as a function of $\WK$ for different bins of $\CT$:
Points: Original data (filled red circles) and background-subtracted data 
(open black circles). Curves: Total fit (solid black), 
\SOneOne contribution (dashed black), integrated background Breit-Wigner (dotted black),
total background (\SOneOne + broad BW, solid blue) and narrow BW (solid red).}
\label{fig:ang}
\end{figure*}

\begin{table}[b]
\centering
\begin{tabular}{|l|c|c|c|}
\hline 
 & $W_R$ & $\Gamma_R$ & $\sqrt{b_{\eta}}A_{1/2}^n$ \\
 & [MeV] & [MeV] & [$10^{-3}$GeV$^{-1/2}$] \\
\hline
Kin.~I   $2\gamma$ & 1670$\pm$1 & 27$\pm$3  (50$\pm$3)	& 12.1$\pm$0.8 \\
Kin.~II  $2\gamma$ & 1669$\pm$1 & 25$\pm$6  (44$\pm$5)	& 11.8$\pm$1.0 \\
Kin.~I   $3\pi^0$  & 1669$\pm$1 & 30$\pm$5  (49$\pm$4)	& 12.9$\pm$0.8 \\
Kin.~II  $3\pi^0$  & 1665$\pm$3 & 53$\pm$17 (66$\pm$14) & 15.6$\pm$2.7 \\
\hline 
Best estimate      & 1670$\pm$5 & 28$\pm$5  (50$\pm$10) & 12.3$\pm$0.8 \\
\hline 
TOF I  $2\gamma$   & 1658$\pm$2 & (42$\pm$4) & 13.2$\pm$0.7 \\
TOF II $2\gamma$   & 1651$\pm$3 & (45$\pm$8) & 18.1$\pm$1.7 \\
TOF I  $3\pi^0$    & 1658$\pm$3 & (41$\pm$9) & 13.9$\pm$1.5 \\
TOF II $3\pi^0$    & 1663$\pm$3 & (20$\pm$9) & 11.3$\pm$2.0 \\
\hline 
Best estimate      & 1658$\pm$7 & (42$\pm$10) & 13.3$\pm$2.0 \\
\hline
\end{tabular}
\caption{Overview of the extracted parameters from the phenomenological fits shown in 
Fig.~\ref{fig:pos_width}: The values in parentheses correspond to the fits where the $W$ 
resolution was not taken into account via convolution with the signal parameterization. 
Uncertainties are statistical only, except for the couplings of the kinematic reconstruction and the
`best estimates' which reflect also the scatter between the different fits and analyses, respectively.}
\label{table:fit_params}
\end{table}

The nature of the narrow structure observed for the $\gamma n\rightarrow n\eta$ reaction
around invariant masses of $W$ = 1670~MeV is not yet understood. The phenomenological
properties of this structure were analyzed with the same simplified ansatz as in 
\cite{Jaegle_11}. It consists of a Breit-Wigner (BW) function with energy-dependent width 
for the contribution of the \SOneOne resonance, a narrow standard BW function for the 
observed structure, and an additional broad BW function parameterizing the remaining 
background contributions at higher energies. The data obtained from the kinematic $W$ 
reconstruction and the reconstruction via TOF (for $\CTTOF$) were fitted separately. 
It was found that the broad BW function was not needed to describe the data from the TOF 
reconstruction (this is so because those data are restricted to $\eta$-backward angles
where the background is much different from forward angles). Individual fits for the 
$\EtaTwoG$ and $\EtaSixG$ data were performed. The results are shown in Fig.~\ref{fig:pos_width} 
and the extracted parameters for position $W_R$, width $\Gamma_R$, and 
electromagnetic coupling $A^n_{1/2}$ (multiplied by the square root of the unknown $N\eta$ 
branching ratio $b_{\eta}$) assuming an $J=1/2$ state are summarized in 
Table \ref{table:fit_params}. Kin.~I, Kin.~II 
and TOF I, TOF II represent datasets obtained with different analysis cuts, where in the 
sets II more strict cuts on the $\eta$ missing mass ($\pm0.5\sigma$), the $\eta$-$n$ 
coplanarity ($\pm0.5\sigma$) and the reconstructed Fermi momentum ($p_F<80$ MeV) were 
applied (see Sec.~\ref{sec:iden}).

The data depending on $\WK$ were additionally analyzed with a fit taking into account the 
resolution of the $W$ reconstruction, which was estimated via simulation 
(see Sec.~\ref{sec:w_rec}). While the extracted parameters for position and coupling 
did not vary much so that they could be simply averaged, the extracted width was 
considerably reduced from around 50 MeV to 30 MeV. This indicates that a significant fraction
of the observed width is related to the experimental resolution and that the intrinsic width is
narrower. The width extracted this way can hence be seen as an approximation of the true width 
while the width obtained with the standard fit corresponds to an upper limit only. 

With the exception of the parameters extracted from the Kin.~II analysis of the $\EtaSixG$ 
channel, which suffer from a large reduction in statistics, all parameters 
corresponding to the kinematic $W$ reconstruction are in good agreement within statistical 
uncertainties. A slight improvement in resolution can be seen for the Kin.~II analysis of the 
$\EtaTwoG$ channel leading to smaller parameters for the width. No such effect can be seen 
in the $\EtaSixG$ data where the reliability of the fit seems to be reduced by lower 
statistics. A best estimate was calculated only taking into account the type I analyses,
as they have better statistics than the type II analyses, and the values are shown in 
Table \ref{table:fit_params}.

\begin{figure}[t]
\centering
\includegraphics[width=0.48\textwidth]{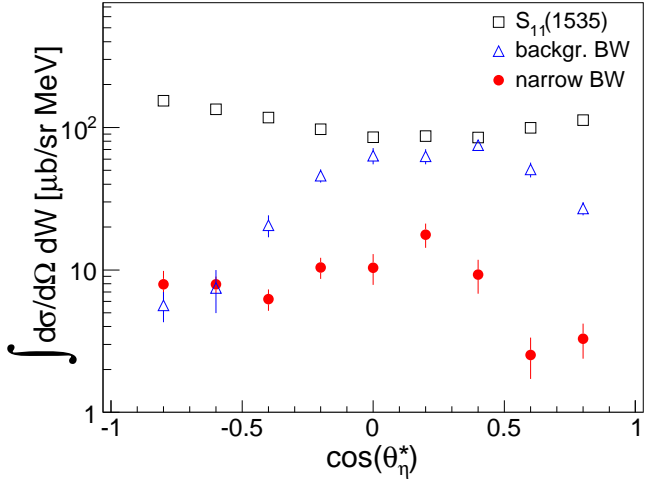}
\caption{(Color online) Contributions from $S_{11}$ BW (open, black squares), background function (open,
blue triangles), and narrow BW (red dots) for different angular bins integrated over
excitation energy. (The most extreme angles have been omitted because of unstable
fit results due to lack of statistics.)}
\label{fig:ang_int}
\end{figure}

The data from the TOF reconstruction cover only $1/4$ of the solid angle resulting in much 
lower statistics. Nevertheless, they serve as an independent check for the presence and the 
properties of the structure. Somewhat lower values for the position and width were obtained.
The coupling was estimated assuming an isotropic angular distribution, which 
(see Figs.~\ref{fig:ang},\ref{fig:ang_int}) is only a rough approximation, but the results 
are in fair agreement with the other analysis.
The smaller width is surprising since the $W$ resolution of the TOF 
reconstruction was estimated to be twice that of the kinematic reconstruction, although 
this is probably too pessimistic. The reason for the different parameter values will be 
discussed below. The fits of the data from the analyses with narrower cuts 
are less reliable due to poorer statistics.
Therefore, as for the kinematic reconstruction method, only the type I data were used to deduce 
a total best estimate of the parameters.

In Fig.~\ref{fig:ang} the differential cross sections are presented as a function of 
$\WK$ for different bins of $\CT$. The same phenomenological fits as discussed above 
were performed to reveal the angular dependence of the structure. The position varies 
between $W_R = 1665$ MeV at backward angles and $W_R = 1680$ MeV at forward angles. 
Also the width is reduced at backward angles. This explains the lower values for position 
and width also observed in the results of the TOF reconstruction where $\CTTOF$. 
On the one hand, the shifting position disfavors the scenario of a single resonance. 
On the other hand, it could also be caused by the simplified
ansatz for the phenomenological fitting (which does of course not include any interference
effects). The angular dependence of the strength of the narrow structure is shown in
Fig.~\ref{fig:ang_int}. Due to the simplified ansatz this is only a qualitative indication
for the variation of the strength over the angular distribution. The figure shows
the contributions of the three fit components integrated over the excitation energy.
The `$S_{11}$' contribution shows the expected behavior (since only one BW function was used
this reflects effectively the contribution from the $S_{11}$ and the $S_{11}$-$D_{13}$
interference, which peaks at forward and backward angles and has a minimum around $\CT = 0$). 
The phenomenological background subsumes contributions from higher lying $P$- and $D$-states,
their interferences, and non-resonant background and has therefore no simple interpretation.
The angular dependence of the narrow structure does not agree with the most simple scenarios 
for its nature, e.g., not with a narrow $P_{11}$ state interfering with the broad $S_{11}$ 
states. The angular distribution of a $P_{11}$ state is isotropic and the interference term 
between $P_{11}$ and $S_{11}$ is proportional to $\CT$. The resulting angular distribution 
would thus have a maximum at forward angles and a minimum at backward angles or vice versa
(depending on the sign of the interference). 
However, Figs.~\ref{fig:ang},\ref{fig:ang_int} show that the structure almost vanishes at
extreme forward angles and is also small at extreme backward angles. Its largest contribution
lies between $\CT\pm 0.5$.
A recent fit of the Bonn-Gatchina partial wave analysis \cite{Anisovich_14} reproduced
the peak-like structure in the total cross section and also the angular distributions in the 
corresponding energy range. In this solution the bump in the total cross section is caused
by interference effects in the $S_{11}$ partial wave. This interpretation requires a sign 
change (relative to the value given by PDG \cite{PDG_12}) of the electromagnetic coupling of 
the $S_{11}$(1650) for the neutron.

\begin{figure*}[t]
\centering
\includegraphics[width=0.99\textwidth]{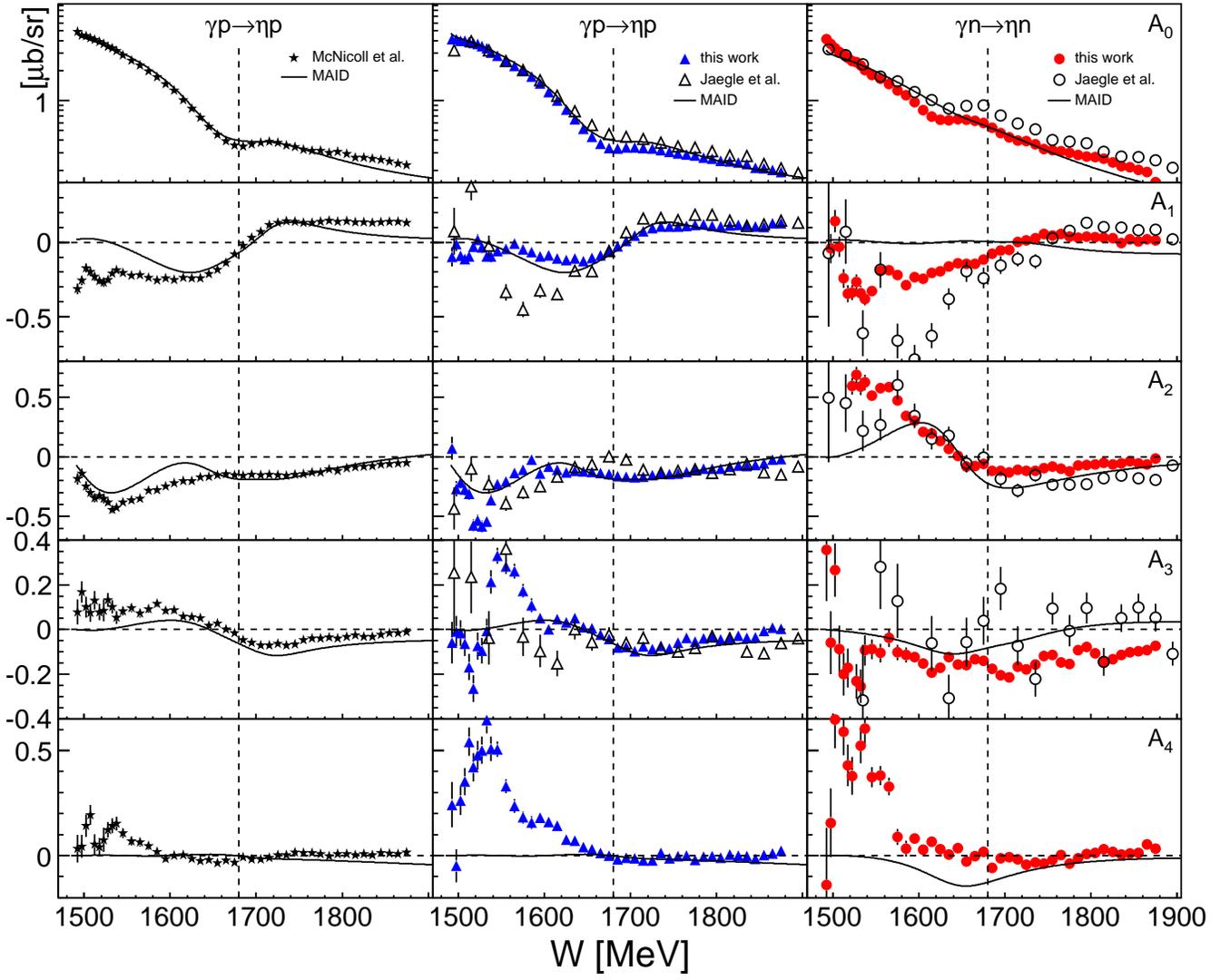}
\caption{(Color online) Comparison of Legendre coefficients $A_i$ extracted from fits of the differential 
cross sections: quasifree proton from this work (filled blue triangles) and 
CBELSA/TAPS \cite{Jaegle_11}  (open black triangles), quasifree neutron from this work 
(filled red circles) and CBELSA/TAPS (open black circles) and free proton data from 
MAMI-C \cite{McNicoll_10} (black stars). Solid lines: MAID model predictions 
\cite{Chiang_02}. Dashed vertical lines: markers at $W = 1680$ MeV.}
\label{fig:legendre}
\end{figure*}

Finally, a comparison of the Legendre coefficients $A_i$ is shown in Fig.~\ref{fig:legendre}.
The $A_i$ were obtained by fitting the angular distributions with a series of Legendre polynomials
$P_i$ up to fourth order
\begin{equation}
\frac{d\sigma}{d\Omega}(W, \CT) = \frac{q_{\eta}^{*}(W)}{k_{\gamma}^{*}(W)}
\sum_{i=0}^{4} A_i(W) P_i(\CT)\,,
\end{equation}
where $q_{\eta}^{*}$ and $k_{\gamma}^{*}$ are the $\eta$ and photon momenta in the 
center-of-mass frame, respectively. The data from the exclusive proton and neutron analyses 
are plotted along with those from CBELSA/TAPS \cite{Jaegle_11} and the free proton 
measurement from MAMI-C \cite{McNicoll_10}. The energy dependences of the proton data of 
this work are in general close to the latter, which indicates 
that cross sections can be reliably extracted from measurements in quasifree kinematics. 
Below $W = 1600$ MeV, some larger discrepancies especially for $A_1$ and $A_3$ are 
observed. This is probably caused by the proton detection efficiency which is 
problematic in this region, as discussed in Sec.~\ref{sec:cs_extr}. Nevertheless, 
above this energy, there is better agreement and also the data from 
CBELSA/TAPS are close to our results. The description of the proton data by the 
MAID model \cite{Chiang_02} is, as expected, reasonable. 

As already seen in the neutron differential cross section data, there are some  
discrepancies between the current results and those from CBELSA/TAPS, although the 
general trends are confirmed. The most significant discrepancy (note
the logarithmic scale) is in the $A_0$ coefficient, which (apart from the phase-space
factor $q_{\eta}^{*}/k_{\gamma}^{*}$) is proportional to the total
cross section. The sign changes in the vicinity of $W = 1680$ MeV of $A_1$ and $A_2$ 
are reproduced. In case of the latter, the different signs for proton and neutron
at low energies are due to interference between the \SOneOne and the \DOneThree resonances
\cite{Weiss_03}. $A_2$ is proportional (neglecting other contributions) to the helicity 
couplings $A^N_{1/2}$  of these states, which have equal signs for protons and neutrons for 
the \DOneThree state, whereas they are opposite for the \SOneOne resonance. This is more 
or less reproduced by MAID, while the model fails  in the description of $A_1$. 
In the discussion of 
the results from $\eta$ electroproduction the change of sign in $A_1$ was interpreted as 
$s$-$p$ wave interference \cite{Denizli_07}. If only $S_{11}$ ($E_{0+}$ multipole) and 
$P_{11}$ ($M_{1-}$ multipole) states are considered, $A_1$ would be directly proportional 
to $\mathrm{Re}(E_{0+}^*M_{1-})$. A change of sign would then mean that the relative phase 
between the two multipoles is changing rapidly due to one of them passing through a resonance. The rough
picture of $A_3$ given by the CBELSA/TAPS measurement is now clarified by the better statistical
quality of the present results. The coefficient seems to be negative throughout the entire energy region.  

\section{Summary and conclusions}
Differential and total cross sections of $\eta$ photoproduction off the proton and 
the neutron were simultaneously measured in quasifree kinematics on a deuteron target 
from threshold up to $E_\gamma = 1.4$ GeV. The $\eta$ mesons were identified using the 
two neutral decays $\EtaTwoG$ and $\EtaSixG$. Exclusive measurements were performed by 
detecting the recoil nucleons.  The total dataset included $4.29\times 10^6$ 
events of inclusive $\eta$ production as well as $1.86\times 10^6$ events with coincident 
protons and $0.63\times 10^6$ events with coincident neutrons. With the inclusive 
measurement, the systematic uncertainties of the nucleon detection efficiencies could be 
checked via $\sigma_{(N)} = \sigma_{p} + \sigma_{n}$ knowing that coherent contributions 
are very small. Effects from Fermi motion were avoided by 
a reconstruction of the center-of-mass energy $W$ from the final state. The technical 
procedure of a kinematic reconstruction of the kinetic energy of the recoil nucleons was 
cross-checked by a time-of-flight measurement. Both methods are only affected by the 
corresponding detector resolution, which for the kinematic reconstruction was determined 
by MC simulations to be $\Delta W < 40$ MeV (FWHM). The results for $\ReacExcP$ 
are mostly in good agreement with data from inclusive hydrogen measurements taking 
into account the poorer resolution in $W$ and effects from the complicated proton efficiency.
The results for $\ReacExcN$ are of unprecedented statistical quality and confirm the 
existence of a peak in the total cross section at $W_R = (1670\pm 5)$ MeV with a width of 
$\Gamma_R = (50\pm 10)$ MeV. Correcting for the finite experimental resolution gave an estimate of 
the intrinsic width of $\Gamma_R = (28\pm 5)$ MeV. If the structure would be related to a
single $J=1/2$ state its strengths would correspond to 
$\sqrt{b_{\eta}}A_{1/2}^n = (12.3\pm 0.8) \times 10^{-3}\,\mathrm{GeV}^{-1/2}$.
However, the precise differential cross sections revealed that the strength depends
on $\CT$; in particular it is suppressed at extreme backward and forward angles which 
disfavors such a scenario. More sophisticated partial-wave analyses of the data are 
under way. First results in the framework of the BnGn model \cite{Anisovich_14} describe 
the data better with a scenario where the main effect is related to interferences in the 
$S_{11}$ sector than with the introduction of a narrow $P_{11}$ state. However, also in 
this approach contributions from other partial waves are needed to reproduce the non-trivial 
angular distributions.

We wish to acknowledge the outstanding support of the accelerator group and operators of 
MAMI. This work was supported by Schweizerischer Nationalfonds (200020-132799, 121781, 
117601, 113511), Deutsche Forschungsgemeinschaft (SFB 443, SFB/TR 16), 
DFG-RFBR (Grant No. 05-02-04014), UK Science and Technology Facilities Council, 
(STFC 57071/1, 50727/1), European Community-Research Infrastructure Activity (FP6), 
the U.S. DOE, U.S. NSF, and NSERC (Canada).

\bibliographystyle{apsrev} 
\bibliography{paper}

\end{document}